\documentclass[
  reprint,
  twocolumn,
  english,
  aps,
  prb,
  superscriptaddress,
  bibnotes,
  amsmath,
  amssymb,
  floatfix
]{revtex4-2}

\usepackage[utf8]{inputenc}
\usepackage[T1]{fontenc}

\usepackage{
  physics, 
  siunitx, 
  graphicx, 
  comment, 
  dcolumn, 
  bm, 
  mathtools, 
  xcolor
}
\usepackage
[colorlinks=true, allcolors=blue]
{hyperref}

\begin{document}
\twocolumngrid

\preprint{APS/123-QED}

\title{Quantum inference on a classically trained quantum extreme learning machine}

\author{Emanuele Brusaschi}
\author{Marco Clementi}
\author{Marco Liscidini}
\author{Matteo Galli}
\affiliation{Dipartimento di Fisica “A. Volta”, Università di Pavia, Via Bassi 6, 27100 Pavia, Italy}
\author{Daniele Bajoni}
\affiliation{Dipartimento di Ingegneria Industriale e dell’Informazione, Università di Pavia, Via Ferrata 1, 27100 Pavia, Italy}
\author{Massimo Borghi}
\email{corresponding author: massimo.borghi@unipv.it}
\affiliation{Dipartimento di Fisica “A. Volta”, Università di Pavia, Via Bassi 6, 27100 Pavia, Italy}

\date{\today}

\begin{abstract}
\noindent
Quantum extreme learning machines (QELMs) are unconventional computing architectures that bear remarkable promise in both classical and quantum machine-learning tasks, such as the estimation of quantum state properties. 
However, the probabilistic nature of quantum measurements  
demands extensive repetitions for training to precisely estimate expectation values, imposing stringent trade-offs among experimental resources, acquisition time, and signal-to-noise ratio, particularly for large datasets.
\textcolor{black}{Here, we introduce a paradigm shift by training the QELM exclusively with intense classical fields, namely twin beams generated by stimulated emission, while performing inference directly on previously unseen genuine quantum input states to predict their quantum properties.} This strategy dramatically reduces acquisition times while substantially enhancing the signal-to-noise ratio. 
Using frequency-bin–encoded biphoton states -- implemented here for the first time in a quantum machine-learning architecture -- we demonstrate entanglement witnessing of two-qubit states with $(93 \pm 4)\%$ accuracy, multi-dimensional entanglement detection, and learning of the Hamiltonian governing photon-pair generation with a fidelity of $(96 \pm 4)\%$.
\textcolor{black}{Our results open a new pathway toward faster and more robust training of photonic quantum extreme learning machines for quantum feature extraction.}
\end{abstract}

\maketitle


\begin{figure*}[t!]    \includegraphics[width=0.95\linewidth]{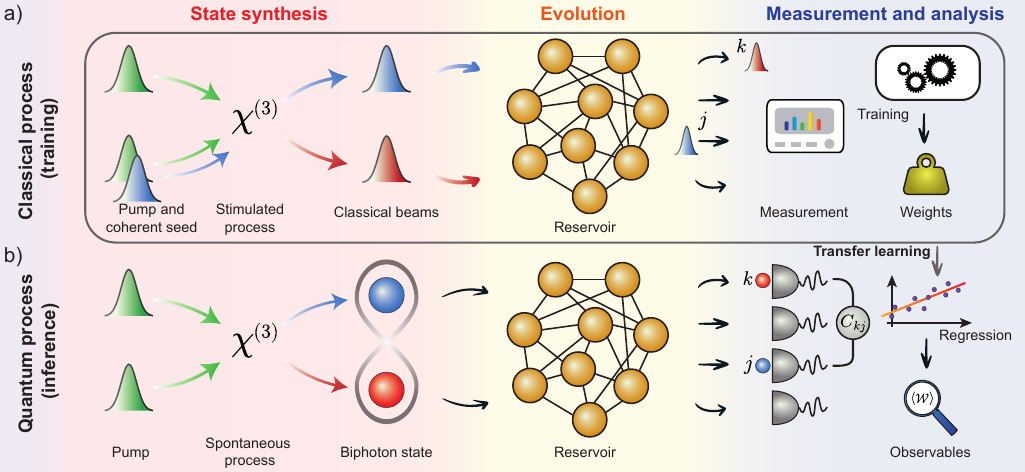}
    \caption{\textbf{Photonic quantum extreme learning machine and its classical training.}
    (a) In the paradigm shift introduced here, the QELM is trained using classical light by exploiting the  correspondence between stimulated and spontaneous emission. A properly amplitude-shaped coherent state (blue) seeds a third-order parametric nonlinear process, such as four-wave mixing. The resulting stimulated coherent beam (red) then propagates through the reservoir (here represented as a network of interconnected nodes without recurrent memory), and its output spectrum is recorded by a spectrum analyzer. If the seed is prepared in one of the asymptotic states of the full system, for example, in spectral mode $j$, the spectral intensity of the stimulated mode $k$ at the output becomes proportional to the photon-pair emission rate in the mode pairs $(k,j)$ in the corresponding spontaneous process. This duality is exploited to train the QELM, thereby determining the weights used in the linear regression. (b) Once the classical training phase is complete, inference is performed on quantum input states generated via the corresponding spontaneous process, such as spontaneous four-wave mixing. This process involves the inelastic scattering of two pump photons, leading to the generation of entangled biphoton states. After generation, the quantum state propagates through the same reservoir used during classical training. The photon-pair detection probabilities are then processed through the linear regression model previously trained with classical signals (transfer learning) to infer the expectation value of a property of the input quantum state, such as the entanglement witness $\langle \mathcal{W}\rangle$. Using classical beams as seeds significantly increases the output photon number compared to coincidence measurements, thereby reducing the time required to train the QELM by several orders of magnitude.
    }
    \label{Fig_1}
\end{figure*}

\section*{\label{sec:Introduction} Introduction}
\noindent 
Quantum Extreme learning Machines (QELMs) are emerging
paradigms in quantum machine learning that aim to
accelerate neural network models by harnessing the fundamental principles of quantum mechanics \cite{innocenti2023potential, xiong2025fundamental}.
Their computational substrate consists of a quantum system composed of interconnected nodes, where the input to be processed can take the form of  information,  encoded into the quantum state of the reservoir nodes \cite{gong2025enhanced,de2025harnessing,joly2025harnessing,rambach2025photonic},  or it may itself be a quantum state interacting with the reservoir \cite{zia2025quantum,suprano2024experimental,assil2025entanglement,vetrano2025state}. \textcolor{black}{In contrast to conventional quantum reservoir computing (QRC) \cite{PhysRevApplied.8.024030,Ghosh2019}, QELMs are memoryless feed-forward architectures that operate on time-independent data. Consequently, while they are not naturally suited for processing temporal information or simulating dynamical systems, they offer significant implementation advantages on noisy hardware.} 
Once injected, the input information spreads across the reservoir nodes, mimicking the layered transformations performed by hidden layers in deep-learning models.   
The states of the output nodes are measured over a set of observables, whose expectation values provide an expanded, high-dimensional evolution of the input data, constituting the output layer of the network. 
Only the weights of this layer are trained to produce the desired output, which may correspond to a quantum observable \cite{krisnanda2025experimental, suprano2024experimental}, an entanglement quantifier \cite{assil2025entanglement, zia2025quantum}, a categorical label \cite{azam2024optically, gong2025enhanced}, or a classical parameter or function \cite{nerenberg2025photon}.
Training is typically carried out via supervised learning, using labeled examples for each input. Because training reduces to a linear regression over the output-layer features, the number of learned parameters remains small, greatly reducing the time, cost, and complexity associated with the training of hidden layers using backpropagation or the parameter-shift rule \cite{crooks2019gradients}. 
This approach also naturally avoids issues such as vanishing gradients \cite{mcclean2018barren}.
In recent years, several demonstrations of QELMs have been reported across multiple physical platforms, including hot atomic vapors \cite{azam2024optically}, superconducting microwave cavities \cite{krisnanda2025experimental}, and nuclear spins \cite{negoro2018machine}.
Parallel to these experimental advances, theoretical work has increasingly focused on understanding their expressivity \cite{xiong2025fundamental, innocenti2023potential}, the role of reservoir topology and quantum information scrambling \cite{vetrano2025state}, and the contribution of quantum resources to enhanced classification and inference performance \cite{nerenberg2025photon}.
On photonics platforms QELMs have recently been implemented with multiple applications including reconstruction of an unknown polarization state \cite{zia2025quantum}, entanglement witnessing \cite{suprano2024experimental,dibartolo2026efficientclassicaltrainingmodelfree} and image classification tasks \cite{joly2025harnessing, rambach2025photonic,gong2025enhanced}.

A universal challenge in QELMs is the need to evaluate expectation values, which requires repeated experimental runs. This repetition is necessary to suppress finite-sampling noise arising from the probabilistic nature of quantum measurements.  
\textcolor{black}{A related challenge is the concentration of observables, namely the tendency of measured observables to become nearly indistinguishable for different input states as the reservoir size increases \cite{xiong2025fundamental}. As a result, an exponentially large number of measurement shots is required to reliably discriminate between them. Consequently, long acquisition times are needed to reduce finite-sampling effects. However, for faint and fragile quantum signals, the required acquisition times may become impractically long and are more susceptible to long-term drifts in the experimental conditions, leading to additional complications, particularly for large training datasets.  
The development of strategies that increase the sampling rate and reduce the acquisition time is therefore highly valuable for mitigating these challenges in both QELM and QRC architectures.}

\textcolor{black}{
In this context, inspiration can be drawn from scenarios in which quantum expectation values have direct classical counterparts, corresponding to macroscopic observables that can be measured efficiently and with high precision.} A well-known example is provided by the Einstein relations, which connect the small rate of spontaneous emission to the much larger rate of stimulated emission in light–matter interactions \cite{Einstein1917}. A similar duality appears in nonlinear optics, particularly between spontaneous parametric processes and parametric amplification in both the low- and high-gain regimes \cite{liscidini2013stimulated,fang2014fast, rozema2015characterizing, eckstein2014high, Grassani2016,triginer2020understanding, gwak2025completely}. At a fundamental level, these general  correspondence principles arise because spontaneous and stimulated processes are governed by the same interaction Hamiltonian, which induces transition rates proportional either to the number of photons in the stimulating field or to vacuum fluctuations, respectively.
\textcolor{black}{Additionally, one can exploit the direct correspondence between the single-photon detection probabilities at the output of a linear optical network and the output intensities of coherent states propagating through the same network. This correspondence is widely used for the characterization of large-scale boson sampling circuits \cite{wang2019boson} and for quantum process tomography \cite{lu2023characterization}.} \\
\begin{figure*}[t!]    \includegraphics[width=\linewidth]{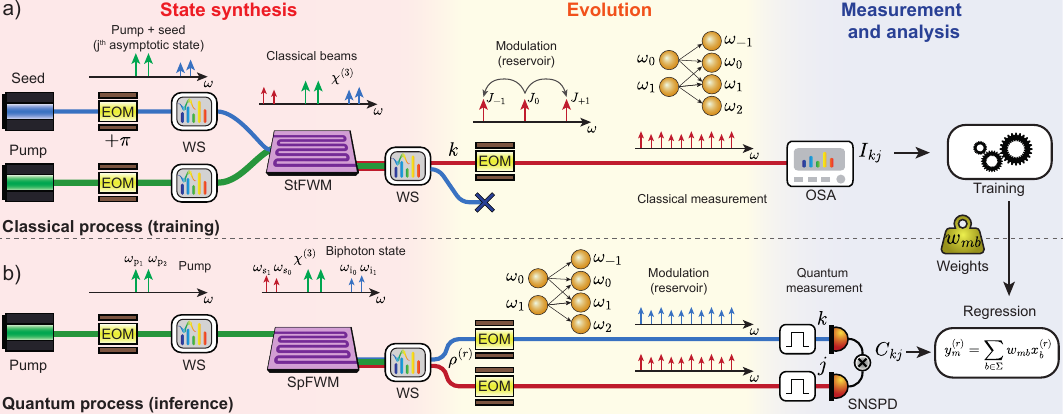}
    \caption{\textbf{Experimental implementation of a frequency-bin QELM.} (a) Training of the QELM is performed by stimulated FWM. To obtain an output intensity of the stimulated beam which is proportional to the coincidence probability $C_{kj}$ of detecting a signal-idler photon-pair in frequency modes $k$ and $j$ in the corresponding spontaneous process, a laser seed (blue) is shaped into a superposition of frequency-bins that mimics the asymptotic output state of the $j^{\mathrm{th}}$ frequency mode of the system (see \cite{liscidini2012asymptotic} and Supplementary Materials).
    The stimulated beam (red) is a coherent state that undergoes evolution through the reservoir, and its spectrum is acquired by an OSA.
    The output intensity $I_{kj}$ on the $k^{\mathrm{th}}$ bin is proportional to $C_{kj}$. The intensities $I_{kj}$ are used to train the QELM using linear regression, from which the weights are computed. (b) Inference is performed on quantum input states generated by SpFWM. Light from a pump laser (green) is sent through a sequence of an EOM and a WS, tailoring its spectrum.
    The pump is routed to an integrated silicon photonic waveguide, where broad-band entangled photon-pairs are generated by SpFWM.
    A second WS completes the state synthesis by defining the output frequency-bin-encoded biphoton state by setting amplitude and phase to the frequency-bins.
    The state evolves through the reservoir, consisting of two EOMs acting on the demultiplexed signal and idler modes, realizing a tight-binding-type interaction, whose intensity and phase depend on the RF driving of the two EOMs.
    The reservoir mixes the frequency-bins, and expands the dimensionality of the Hilbert space.  
    After narrowband filtering, which selects one of the possible signal-idler frequency-bins combinations $(k,j)$, coincidence detection is performed by a pair of SNSPDs, yielding the value $C_{kj}$. The multiple frequency-bin combinations represent the output layer of the QELM. These are used as inputs to the linear regression, which is applied using the weights assessed in the training phase. 
    EOM: electro-optic phase modulator. 
    WS: waveshaper.
    SpFWM: spontaneous four-wave mixing.
    StFWM: stimulated four-wave mixing.
    SNSPD: superconducting nanowire single-photon detector.
    OSA: optical spectrum analyzer.}
    \label{Fig_2}
\end{figure*}
\textcolor{black}{Inspired by these analogies, we introduce and experimentally demonstrate the new paradigm of  classical training of quantum extreme learning machines. Specifically, we train the machine using only strong classical signals, namely twin beams generated by stimulated emission, and subsequently demonstrate its inference capabilities on quantum input states.} 
We show that this \emph{classical training} enables excellent inference performance of nonclassical observables on quantum inputs, such as entanglement \cite{zia2025quantum, Ghosh2019}, high-dimensional entanglement quantifiers \cite{wang2018multidimensional}, and properties of quantum states and process Hamiltonians.
Compared with training based on conventional coincidence measurements, such classical operation dramatically reduces the training time, while improving at the same time the signal-to-noise ratio (SNR).
Experimentally, we leverage frequency-bins \cite{Lu2023} to encode biphoton states generated via spontaneous four-wave mixing (SpFWM) on a photonic chip \cite{kues2017chip, clementi2023programmable, borghi2023reconfigurable}.
This experimental demonstration naturally benefits from the ease of scaling towards high dimensions, here showcased up to ququarts, and represents the first implementation of a quantum learning machine based on frequency-bins.
Electro-optic modulators (EOMs), driven by radio-frequency signals matching the bin spacing, constitute the reservoir, scrambling qubits and qudits to higher-dimensional (up to 64) Hilbert spaces.
The set of all possible frequency correlations between the signal and idler photons forms the output layer of the QELM.
We benchmark the quantum inference of the classically trained machine with quantum light against expected outcomes, demonstrating strong qualitative and quantitative performance across all cases. 
\textcolor{black}{Although experimentally demonstrated here for two qudits, we theoretically prove in the Supplementary Materials that the classical training protocol can be generalized to an arbitrary number of parties and dimensions. We further show that it extends to the continuous-variable regime, enabling the inference of first- and second-order moments of arbitrary input states.}

\section*{\label{sec:Results} Results}
\subsection*{Overview and formalism}

\noindent
Generally, the QELM workflow consists of three sequential steps: input-state preparation, evolution through the reservoir system, and measurement of the network nodes. The reservoir evolution can be described by a map that sends the initial state in the joint input–reservoir space to the final state of the reservoir  \cite{innocenti2023potential}.
In practice, a strict distinction between the degrees of freedom of the input and those of the reservoir is not required, as the input state can be initially embedded into a subset of modes forming a high-dimensional qudit that also hosts the reservoir modes (typically initialized in the vacuum state).
Moreover, the measurements may also involve the state degrees of freedom \cite{innocenti2023potential}. This is the framework that we adopt throughout this work.
After the evolution, a set of measurements, described by positive-operator-valued measures (POVMs) $\{\mu_b:b\in \Sigma\}$, where $\Sigma$ is the number of measurement outcomes, is performed on each state $r$. This procedure yields the probabilities $x_{b}^{(r)}$ for all the input states of the training set. 

QELM training is a supervised learning process that associates each input training state $\rho^{(r)}$ with corresponding $M$ labels or values $\{y^{(r)}_m\}_{m=0}^{M-1}$ (e.g., an entanglement measure or a set of density matrix elements), and the goal is to determine the set of weights $w_{mb}$ that best approximates the linear function
\begin{equation}
    y^{(r)}_m \sim \sum_{b\in\Sigma}w_{mb} x_{b}^{(r)} = \sum_{b \in \Sigma} w_{mb} \textrm{Tr}(\mu_b \rho^{(r)}), \label{eq:QELM_learning}
\end{equation}
which represents a standard linear regression problem, for which closed-form solutions exist for the weights $w_{mb}$ (see Methods for more details). Once the weights have been determined, the inference accuracy is evaluated on a set of $N_\textrm{t}$ previously unseen states, using Eq.~(\ref{eq:QELM_learning}) to infer the labels $\tilde{y}^{(r)}_m$. The accuracy is quantified using a suitable metric, such as the mean squared error $\textrm{MSE}=(MN_\textrm{t})^{-1}\sum_{rm}|\tilde{y}^{(r)}_m-y^{(r)}_m|^2$. 

The definitions introduced so far illustrate the standard workflow employed in the training and testing of any QELM architecture \cite{innocenti2023potential, mujal2021opportunities, ghosh2021quantumReview}. 
Since the output typically requires the estimation of expectation values, possibly corresponding to observables that carry non-classical correlations arising from entanglement, the stochastic nature of quantum measurements requires the preparation of multiple copies of the state and repeated measurements to accumulate sufficient statistics and suppress finite-sampling noise. 
Although strategies based on ensemble computing have been proposed to reduce the integration time \cite{mujal2021opportunities, negoro2018machine}, the achievable speedup is modest, and they require the implementation of global unitaries acting identically on many copies of identically prepared input states.
\textcolor{black}{The paradigm shift introduced here lies in exploiting the connection between the intensity of a macroscopic classical signal and the outcome probabilities of single photons and correlated photon pairs generated through the SpFWM process and propagating in a linear optical network.  While the former can be readily reproduced by a single coherent beam \cite{laing2012super}, mimicking two-photon correlations requires twin beams produced via stimulated four-wave mixing (StFWM)} \cite{liscidini2013stimulated,azzini2012classical}.
This last concept is illustrated in Fig.~\ref{Fig_1}(a). During the training phase, 
rather than collecting coincidence counts from the weak spontaneous signal to evaluate the coincidence probability $C_{kj}$ between a signal photon in mode $k$ and an idler photon in mode $j$ at the reservoir output, one can detect an informationally equivalent but much stronger field by stimulating the generation of a coherent signal beam, thereby drastically reducing the integration time while simultaneously increasing the SNR. This is achieved by seeding the photon-pair source with a properly \emph{amplitude-shaped} idler beam and by measuring the signal intensity in mode $k$ using a standard detector.
The required amplitude profile of the idler beam can be formally obtained through a simple prescription \cite{borghi2020phase}.
One first considers the corresponding spontaneous process and then back-propagates the path of the idler photon in mode $j$ from the detector to the source. 
The resulting complex amplitude pattern, which is generally multimode and termed the asymptotic output field of the $j^{\mathrm{th}}$ mode of the system \cite{liscidini2012asymptotic}, is 
subsequently used as a coherent seed for stimulated four-wave mixing. 
This process generates a signal beam whose intensity $I_{kj}$ in mode $k$ becomes proportional to $C_{kj}$, thereby mapping the quantum correlations onto a macroscopic classical signal.\\
More rigorously, the coincidence probability $C_{kj}$ at the output of the QELM in the spontaneous process can be written as (see Supplementary Materials for derivation)
\begin{equation}
    C_{kj} \propto |[\mathbf{U}_{\textup{QELM,s}} \mathbf{S}_{\textup{si}}\mathbf{U}_{\textup{QELM,i}}^{T}]_{kj}|^2, \label{eq:coinc_prob}
\end{equation}
where $[\cdot]_{kj}$ denotes the matrix element, $[\mathbf{S}_{\textup{si}}]_{kj}$ is the probability amplitude for the signal and idler photons to be in modes $k$ and $j$, respectively, at the QELM input, and $\mathbf{U}_{\textup{QELM,s(i)}}$ are the matrices describing the transformation implemented by the QELM to the signal and idler photons (here and in the following assumed to factor into independent transformations).
In the stimulated case, by contrast, one starts with the idler seed in a coherent state in mode $j$, described by the amplitude vector $\bm{\alpha}_j=(0,\ldots,\alpha_j,\ldots,0)$, and shapes its amplitude by applying the transformation $\bm{\alpha}_j\mapsto\tilde{\mathbf{U}}_i\bm{\alpha}_j$ before entering the nonlinear source. The intensity $I_{kj}$ of the stimulated coherent state of the signal in mode $k$ is given by (see the Supplementary Materials for the derivation)
\begin{equation}
I_{kj}\propto |\alpha_j|^2 |  [\mathbf{U}_{\textup{QELM,s}}\mathbf{S}_{\textup{si}}\tilde{\mathbf{U}}_{\textup{i}}^*]_{kj}|^2. \label{eq:coherent_intensity}
\end{equation}
Therefore, $C_{kj}\propto I_{kj}$ only if $\tilde{\mathbf{U}}_\textup{i}=\mathbf{U}_{\textup{QELM,i}}^{\dagger}$, i.e., the complex amplitude $\mathbf{U}_{\textup{QELM,i}}^{\dagger}\bm{\alpha}_j$ equals that of a coherent beam back-propagated from mode $j$ at the detectors towards the nonlinear source.\\
The set of intensities $I_{b=kj}$ replaces the POVM outcomes $x_b^{(r)}$ in the spontaneous process, and are associated with the labels $y_m^{(r)}$. Through Eq.(\ref{eq:QELM_learning}), they are used to train the QELM and learn the weights $w_{mb}$.\\
Inference is carried out on previously unseen quantum input states (Fig.\ref{Fig_1}(b)) generated via SpFWM. It is essential that the inference stage operates directly on quantum states, as these constitute the actual quantum resources employed in experimental protocols.

\subsection*{\label{subsec:QELM_implementation} 
Implementing a QELM with frequency-bins}

\noindent
The inference performance of any machine-learning protocol depends critically on the quality of the training dataset, which should be as unbiased as possible, representative of the process under investigation, and of sufficient size to fully capture its complexity. 
Moreover, while the state evolution may be uncalibrated, it must be stable over time and repeatable across multiple runs of the experiment.
These requirements call for a platform capable of preparing a large variety of quantum states with high fidelity, together with a robust and reliable system for their evolution. 
We now outline how all these steps are implemented in our work. We first discuss \emph{state synthesis}, which is accomplished by generating photon-pairs in reconfigurable states via SpFWM in a Silicon-On-Insulator spiral waveguide. 
Logical states are encoded in the discretized frequency degree of freedom (\textit{frequency-bin}), which is particularly attractive for implementing QELMs due to its inherently high-dimensional state space \cite{kues2017chip, imany2019high} and robustness against environmental noise \cite{borghi2023reconfigurable}, and ease of manipulation \cite{imany2020probing, lupo2021photonic, senanian2023programmable}. Next, we discuss how state \emph{evolution} is implemented using an electro-optic phase modulator, which coherently mixes frequency bins and broadcasts the input states into a higher-dimensional Hilbert space. Finally, we present the \emph{measurement} procedure applied to the reservoir output nodes. At this stage, we highlight the main difference between two workflows: the standard approach, in which both QELM training and inference are performed on quantum input states, and the classical-training approach, where training is carried out using stimulated emission and the learned weights are subsequently transferred to the inference stage operating on quantum inputs.
\paragraph*{State synthesis.} 
The quantum states are synthesized as shown in Fig.~\ref{Fig_2}(b). 
A pump laser is phase-modulated at $20$ GHz to generate an electro-optic comb, which then passes through a waveshaper (WS). 

Two coherent pump lines $\omega_{\mathrm{p_1}}$ and $\omega_{\mathrm{p_2}}$ of equal intensity are selected by the WS (dual pump (DP) configuration), which are then amplified and injected into a silicon waveguide.
Broad-band SpFWM occurs in the waveguide, from which we carve discrete frequency-bins with spacing and width of $20$ GHz using the WS. 
These bins are symmetrically positioned in the signal and idler bands at frequencies $\omega_{\mathrm{s}}$ and $\omega_{\mathrm{i}}$, respectively, so as to satisfy energy conservation. 
In the waveguide, either non-degenerate or degenerate SpFWM processes can occur. 
The degenerate processes $2\omega_{\mathrm{p_1}}= \omega_{\mathrm{i_0}}+ \omega_{\mathrm{s_1}}$ and $2\omega_{\mathrm{p_2}}= \omega_{\mathrm{i_1}}+ \omega_{\mathrm{s_0}}$ generate photon-pairs in the $\ket{10}$ and $\ket{01}$ components of the two-qubit state, respectively. 
In contrast, the non-degenerate process $\omega_{\mathrm{p_1}}+\omega_{\mathrm{p_2}}=\omega_{\mathrm{i_0}}+\omega_{\mathrm{s_0}}=\omega_{\mathrm{i_1}}+\omega_{\mathrm{s_1}}$ contributes to the $\ket{00}$ and $\ket{11}$ components of the state.
The WS implements adjustable weights $\mathbf{g}=(g_{\mathrm{s_0}},g_{\mathrm{s_1}},g_{\mathrm{i_0}},g_{\mathrm{i_1}})$ to two selected signal and idler bins $\omega_{\mathrm{\mathrm{s(i)}}{0(1)}}$, while blocking all other frequency components, thereby shaping the quantum state into the form:
\begin{equation}
\begin{aligned}
\ket{\Psi}_{\textup{DP}} 
&= \frac{1}{\mathcal{N}_{\textup{DP}}(\mathbf{g})}
\Bigl(
2 g_{\mathrm{s_0}} g_{\mathrm{i_0}} \ket{00}
+ 2 g_{\mathrm{s_1}} g_{\mathrm{i_1}} \ket{11} \\
&\qquad\qquad
+ g_{\mathrm{s_0}} g_{\mathrm{i_1}} \ket{01}
+ g_{\mathrm{s_1}} g_{\mathrm{i_0}} \ket{10}
\Bigr),
\end{aligned}
\label{eq:DP_states}
\end{equation}
where $\mathcal{N}_{\textup{DP}}(\mathbf{g})$ is a normalization factor and the factor of two arises from the non-degenerate SpFWM process. 
In addition, the WS can be configured to transmit only a single pump line at frequency $\omega_{\mathrm{p}}=(\omega_{\mathrm{p}_1}+\omega_{\mathrm{p}_2})/2$ (we refer to this configuration as single-pump (SP)), extending the class of states that can be synthesized from our setup (see Methods section for more details).

The separable states $\ket{00}$ and $\ket{11}$ can be obtained by setting either $g_{\mathrm{s_1(i_1)}}$ or $g_{\mathrm{s_0(i_0)}}$ to zero in Eq.~(\ref{eq:SP_states}). 
Similarly, separable states in which one of the two photons is in a superposition can be generated in the DP configuration by setting one of the weights to zero. 
For example, by choosing $g_{\mathrm{s_0}}=0$, one obtains the state $\frac{1}{\mathcal{N}_{\textup{DP}}(\mathbf{g})} \ket{1}_\mathrm{s}(g_{\mathrm{i_0}}\ket{0}_\mathrm{i}+2g_{\mathrm{i_1}}\ket{1}_\mathrm{i})$.
Overall, by combining the two pump configurations and adjusting the WS weights, one can access a wide range of entangled and separable states without physically changing the source.

\paragraph*{Evolution.} 
The reservoir is implemented using  electro-optic phase modulators placed along both the signal and idler paths and driven by a sinusoidal waveform at $20$ GHz with modulation depth $\delta=1.4$. 
With a single driving tone, the EOMs implement a tight-binding Hamiltonian of the form
$\mathbf{H}_\mathrm{EOM}=\sum_{m}\frac{\Omega(\delta)}{2}a_m^{\dagger}a_{m-1} + H.c.$,
where $|\Omega(\delta)|^2$ represents the hopping probability between adjacent frequency modes $m$ and $m-1$ (described by the ladder operators $a_m$ and $a_{m-1}$), \cite{haldar2022steering} and can be tuned by adjusting the modulation depth $\delta$ \cite{imany2020probing} (see Supplementary Materials for additional details on the EOM transformation). 
The reservoir topology can be visualized as a linear chain of frequency modes with nearest-neighbor interactions, where  the input states constitute single excitations that are initially localized at specific lattice sites and subsequently spread across the reservoir nodes during the evolution.
In principle, more complex reservoir topologies can be realized by introducing additional driving frequencies in the EOM. 
For example, a two-dimensional square lattice of size $L$ can be implemented in a nonlocal one-dimensional lattice by coupling nearest neighbors and $L^{\textrm{th}}$ nearest neighbors using driving frequencies $\nu$ and $L\nu$, respectively \cite{senanian2023programmable}. 
The frequency domain thus provides a highly versatile platform for exploring a wide variety of synthetic lattices for QELMs, whose performance in terms of accuracy and stability generally improves with increasing network connectivity \cite{innocenti2023potential}.

\paragraph*{Measurement.}
At the output of the EOMs, the reservoir nodes constituted by the frequency-bins are measured pairwise by performing frequency-resolved coincidence detections between the signal and idler modes using superconducting single-photon detectors in combination with narrowband filters.
We consider $Q$ bins for each photon, yielding a $Q\times Q$ grid of frequency correlations, which is vectorized to form the input feature vector $\mathbf{x}^{(r)}$ associated with each input state $\rho^{(r)}$. The training feature vectors are used to determine the regression weights, which are subsequently applied during the inference stage on the test set via Eq. (\ref{eq:QELM_learning}). In the standard workflow, the feature vectors for both the training and test sets are obtained through coincidence measurements. However, the large size of the training set, together with the inherently weak nature of the spontaneous process, requires long integration times to accumulate sufficient statistics for a reliable estimation of the weights in Eq. (\ref{eq:QELM_learning}).
By contrast, in our approach we exploit the connection between the spontaneous and stimulated processes—mathematically expressed by Eqs. (\ref{eq:coinc_prob}, \ref{eq:coherent_intensity})—to extract the feature vectors of the large training set using intensity-only measurements (\emph{classical training}). This strategy enables a drastic reduction of the time dedicated to QELM training.

\subsection*{{\label{subsec:QELM_training}}Classical training}

\begin{figure*}[t!]    \includegraphics[width=\linewidth]{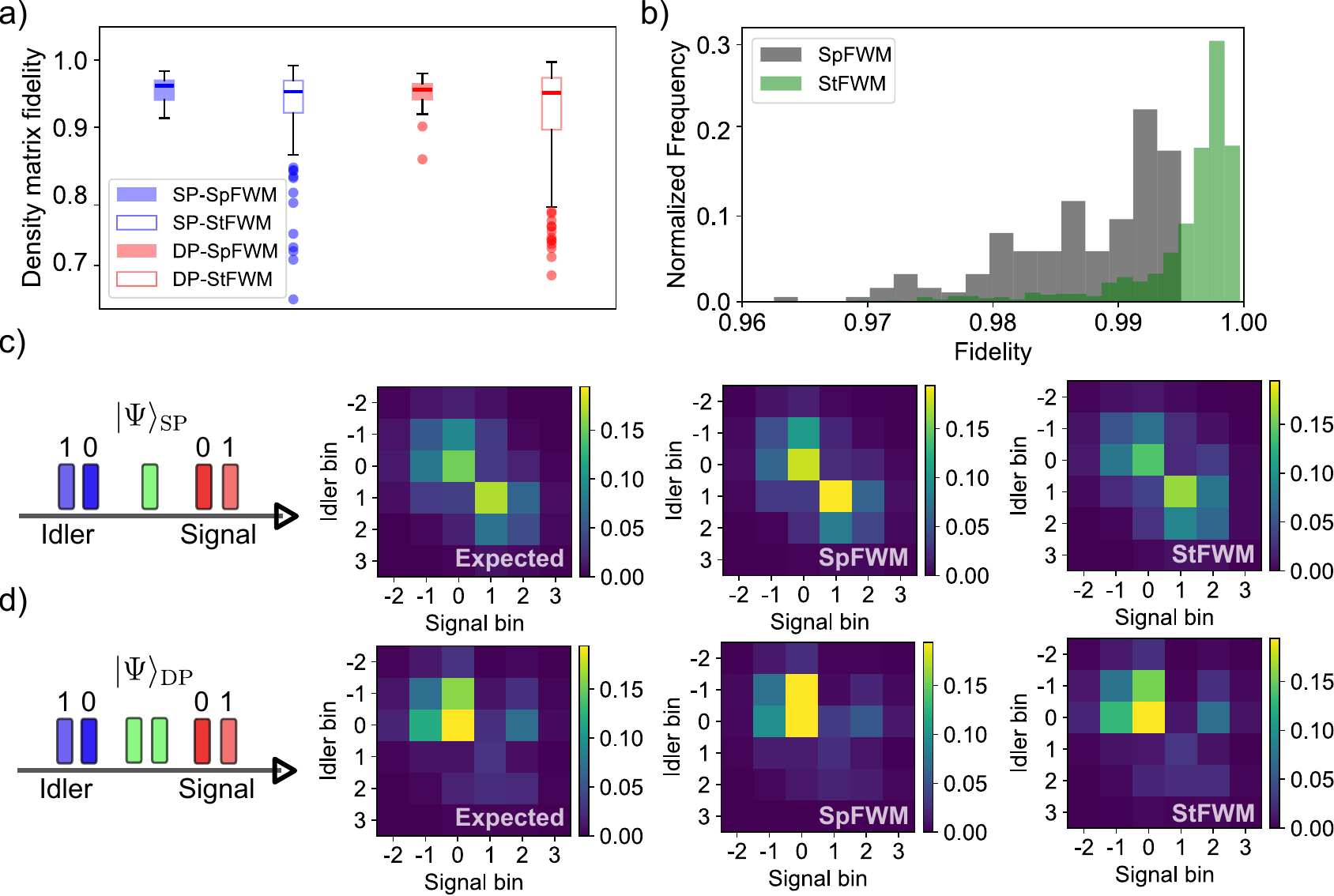}
    \caption{\textbf{Synthesis and characterization of two-qubit states.}
    a) Box-plots showing the fidelity of the reconstructed density matrices with respect to the expected targets. Random states are prepared in the SP (Eq.(\ref{eq:SP_states})) and DP Eq.(\ref{eq:DP_states}) configurations. Datasets are acquired either using StFWM or coincidence measurements (SpFWM). (b) Distribution of the fidelity between the expected $6\times6$ matrices describing the frequency correlations at the output of the QELM and the corresponding experimental patterns acquired via coincidence measurements (SpFWM) and stimulated emission (StFWM).
(c) A representative example of frequency correlations at the output of the QELM for a state prepared in the SP configuration. From left to right: the expected pattern, the pattern obtained via coincidence measurements of spontaneously emitted photon-pairs (SpFWM), and the pattern obtained via StFWM by measuring light intensities with an optical spectrum analyzer. (d) Representative frequency correlation matrix for a state prepared in the DP configuration.} 
    \label{Fig_3}
\end{figure*}
\noindent
The experimental procedure used to train the frequency-bin QELM is illustrated in Fig.~\ref{Fig_2}(a). 
A seed (idler) laser tuned to frequency-bin $j$ passes through an EOM and a WS in series before being coupled into the chip together with the pump beam. The EOM–WS sequence shapes the seed spectral amplitude to match that of the idler photon back-propagated from detector mode $j$ to the source in the spontaneous process shown in Fig.~\ref{Fig_2}(b).
Conjugation (i.e. time-reversal) is implemented by shifting the RF waveform driving the EOM by $\pi$ and by replacing the set of idler weights $\mathbf{g}_\mathrm{i}$ applied by the WS in the spontaneous process with their complex conjugates $\mathbf{g}_\mathrm{i}^*$ (see Supplementary Materials for a rigorous derivation).
Stimulated FWM occurs in the waveguide, generating a signal beam that subsequently passes through the same EOM reservoir shown in Fig.~\ref{Fig_2}(b). An optical spectrum analyzer (OSA) placed at the EOM output is then used to resolve the individual frequency-bins composing the signal beam.
This procedure, which can be naturally extended to higher-dimensional two-qudit systems, is particularly straightforward to implement with frequency-bin encoding, since all modes propagate and are coherently manipulated within the same optical fiber. 
By contrast, other forms of high-dimensional encoding, such as path, time-bin, or orbital angular momentum, may require sub-wavelength stabilization of relative optical paths to preserve coherence.

As a first preliminary step, we evaluated the synthesis capability of our apparatus.  
We randomly sampled the complex weights $\mathbf{g}$ and used them to construct and generate $250$ states in the SP configuration and $500$ states in the DP configuration. \textcolor{black}{Separable states are also included in this set and are obtained by setting either $g_{{\textrm{s}_1}(\textrm{i}_1)}$ or $g_{{\textrm{s}_0}(\textrm{i}_0)}$ to zero. In this case, the training beams are still generated by stimulated emission, but the intensity distribution at the QELM output coincides with that obtained from independent coherent states, each exciting one of the signal and idler modes. Therefore, stimulated emission, combined with the subsequent manipulation through a WS, enables the QELM to be trained with classical fields that mimic both independent single-photon probabilities and two-photon correlations.}
Full quantum state tomography was performed on these ensembles at the output of the synthesis stage using the classical seed \cite{rozema2015characterizing}, yielding Uhlmann fidelities of $0.94(5)$ and $0.92(7)$ with respect to the expected density matrices for the SP and DP configurations, respectively (SP-StFWM and DP-StFWM in Fig.\ref{Fig_3}(a), see also  Supplementary Materials).
We then randomly selected 20 elements from these ensembles for both SP and DP configurations, and performed quantum state tomography via coincidence measurements (SP-SpFWM and DP-SpFWM panels on Fig.\ref{Fig_3}(a)), achieving fidelities of $0.98(2)$ and $0.95(2)$ respectively.\\ 
As a second step, we fed the reservoir using both classical input and quantum states.
The purpose of this test was to assess the ability of a classical process to reproduce the quantum correlations captured by frequency-resolved coincidence detection.
Such correlations will be later used to train the QELM by estimating the weights reported in Eq.~\eqref{eq:QELM_learning}.
To this end, we measured the $Q\times Q$ matrices of frequency-bin correlations at the output of the QELM, setting $Q=6$ and using the bins $\omega_{\mathrm{s(i)}0},\omega_{\mathrm{s(i)}1}$ to define the logical state of each qubit. 
\begin{figure}[t!]    \includegraphics[width=\linewidth]{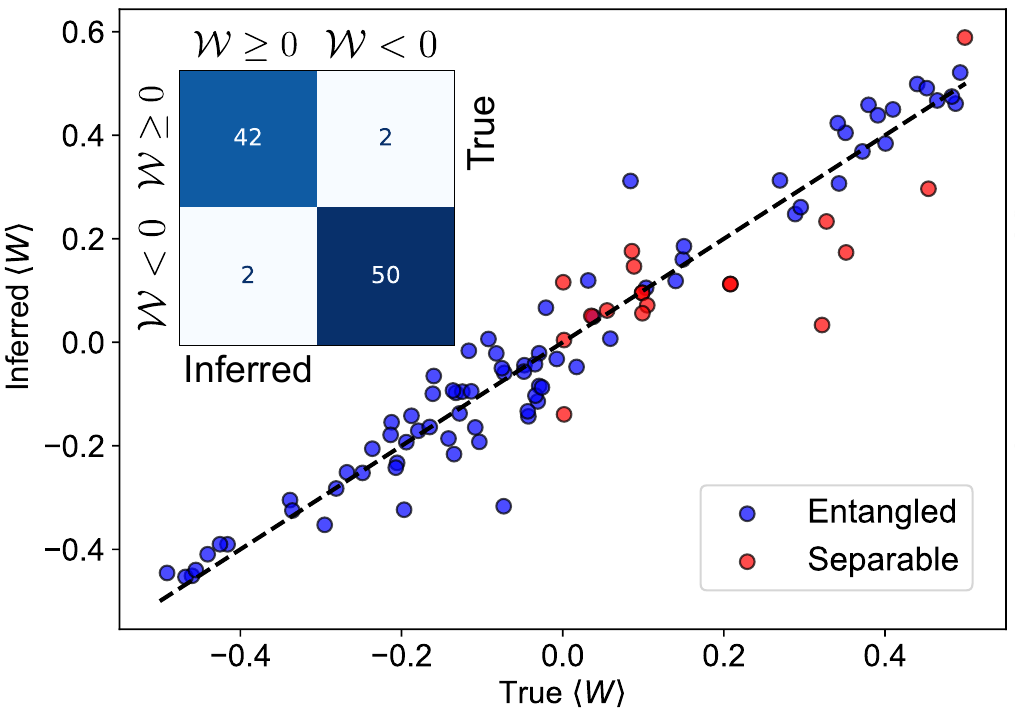}
    \caption{\textbf{Entanglement witness for two qubits.}
Inferred versus true values of the entanglement witness $\langle \mathcal{W} \rangle$ on a representative test set acquired through coincidence measurements of spontaneously emitted photon-pairs. Diamond markers indicate separable states.
The confusion matrix in the inset illustrates the binary classification of states with $\langle \mathcal{W}\rangle \le 0$ and $\langle \mathcal{W}\rangle \ge 0$.  
    }
    \label{Fig_4}
\end{figure}
Leveraging the classical process, we collected $250$ correlation matrices in the SP configuration and $500$ matrices in the DP configuration (including $100$ obtained from separable states). 
Using coincidence measurements, we collected $100$ matrices for both the SP and DP configurations (of which $20$ corresponded to separable states). 
As a comparison, the total acquisition time was approximately $1.5$ hours for completing the stimulated measurements and $24$ hours for the coincidence measurements, the former mainly determined by the OSA acquisition time, the second one by the integration time needed to collect sufficient statistics.
Thus, the use of the stimulated process reduced the integration time per state by a factor of $\sim 60$. 
Remarkably, this reduction in integration time was accompanied by an improvement in the SNR of the brightest frequency-bin correlations of approximately $19$ dB (see Supplementary Materials for a detailed analysis of the SNR).
This result is particularly notable given the very different nature of the two detection processes, which differ in terms of mechanism, cost, complexity, and SNR, the latter in particular being nearly 6 orders of magnitude lower for the standard OSA.
With the use of specialized equipment, such as single-photon OSA \cite{Guidry2022}, the integration time or SNR could be further improved by several orders of magnitude, albeit at the cost of additional complexity.
In Fig.~\ref{Fig_3}(b), we quantitatively compare the distributions of the fidelities (Frobenius norm) between the expected outcomes and the patterns obtained via stimulated emission (StFWM) and coincidence detections (SpFWM). 
On average, these fidelities are  $\mathcal{F}_{\textup{SpFWM}}=0.987(7)$ and $\mathcal{F}_{\textup{StFWM}}=0.995(5)$, respectively. 
Representative examples of patterns associated with SP and DP states, acquired using both stimulated and coincidence measurements, are shown in Fig.~\ref{Fig_3}(c,d), demonstrating excellent agreement with expected outcomes.

\subsection*{\label{subsec:Entanglement_witness_qubits}Entanglement witness of two qubits}
\begin{figure*}[t!]    \includegraphics[width=\linewidth]{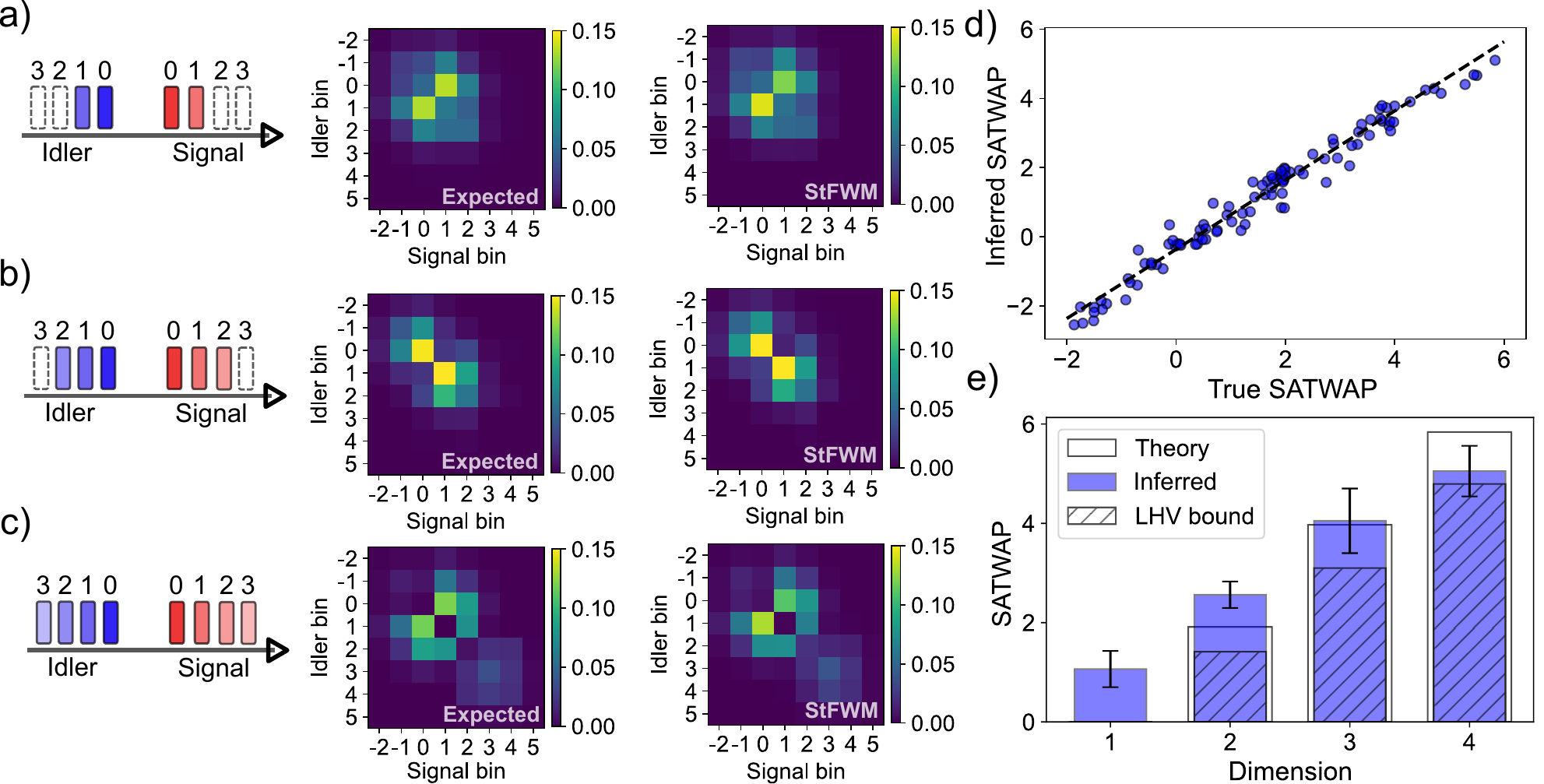}
    \caption{\textbf{Entanglement witness in high-dimensions.}
    (a-c) Representative examples of frequency-bin correlation matrices at the output of the QELM. The panels show the expected patterns (Expected) for randomly chosen qubit (a), qutrit (b), and ququart (c) states. The rightmost column panels display the corresponding patterns measured via stimulated emission (StFWM).
(d) Inferred versus true values of the SATWAP inequality for a representative test set acquired using stimulated emission.
(e) Inferred values (cyan) of the SATWAP inequality for maximally entangled states of dimensions 2, 3, and 4, as well as for the separable state $\ket{00}$ (indicated as dimension one). The corresponding QELM output patterns were acquired via coincidence measurements on spontaneously emitted photon-pairs, while training was performed using stimulated emission. The hatched bars indicate the LHV bound. Error bars are computed from inference obtained by training the QELM using 30 different training sets.   
    \label{Fig_5}}
\end{figure*}

\noindent
The correlation matrices acquired by the stimulated process, as discussed in the previous paragraph, were used to train the QELM to reproduce the value of the entanglement witness: 
\begin{equation}
\mathcal{W}=\frac{I}{2}-\ketbra{\Phi^{+}}{\Phi^{+}},  
\end{equation}
\\
where $\ket{\Phi^+}=\frac{1}{\sqrt{2}}(\ket{00}+\ket{11})$  (see Methods for details on the training procedure). 
Values of $\mathcal{W}<0$ indicate entanglement, while separable states all have $\mathcal{W}\ge0$. 
The trained QELM is used to infer $\mathcal{W}$ on test sets of previously unseen quantum states, whose correlation matrices have been acquired through coincidence measurements.   

A representative example of inferred $\mathcal{W}$ versus its true value is shown in Fig.~\ref{Fig_4}, indicating a very good inference performance. 
We assessed an MSE on the test set of $0.012(7)$ (normalized MSE: $\textrm{NMSE}=0.18(4)$) by averaging over $30$ random compositions of both the training and the test sets. 
From the perspective of certifying entanglement, we quantified the fraction of entangled states with $\mathcal{W}<0$ for which the QELM also infers $\mathcal{W}<0$, showing the confusion matrix related to the inference in Fig.~\ref{Fig_4}. 
The accuracy in entanglement certification is $96\%$, and only slightly decreases to $93(4)\%$ when it is averaged over $30$ independent and randomly populated training and test sets. 
Similarly, separable states erroneously classified as entangled ($\mathcal{W}<0$) are $12(8)\%$. 

\subsection*{High dimensional states}
\noindent
We now generalize entanglement detection to higher dimensions, a task that is particularly suitable to be implemented with frequency-bins, owing to the inherently high dimension of the associated Hilbert space  
\cite{lu2022bayesian, borghi2023reconfigurable, imany2019high}. 
We generate a wide class of two-qudit states of the form $\ket{\Psi}=\sum_{j=0}^{3} \alpha_j \ket{jj}$ with the SP configuration, where the complex coefficients $\alpha_j$ are set by the WS. 

Tests of nonlocality tailored to maximally entangled two-qudit states have been developed to reveal correlations incompatible with local-hidden-variable (LHV) theories.
We choose the Salavrakos–Augusiak–Tura–Wittek–Acín–Pironio (SATWAP) inequality, previously used to certify high-dimensional entanglement in path-encoded qubits \cite{wang2018multidimensional}, as the target feature to be reproduced by our QELM. 
The SATWAP inequality takes the form $I_d\le C_d$, where $I_d$ is a correlator constructed from the joint detection probabilities between the signal and the idler photons, and $C_d$ is the classical bound implied by LHV theories for $d$-dimensional systems (see Methods for details). 
In general, $I_d$ can be constructed from $m\ge2$ measurement settings with $d$ outcomes; larger $m$ values provide increased robustness to depolarizing noise and enable stronger violations of the classical bound \cite{salavrakos2017bell}. 
For this proof-of-concept demonstration we choose $m=2$. 
However, we emphasize that a QELM could, in principle, infer $I_d$ for arbitrary $m$ with a single measurement setting, provided that the reservoir is sufficiently expressive. 
Additionally, the QELM learns the projectors appearing in $I_d$, which correspond to superpositions of multiple frequency-bins with highly specific amplitude profiles -- states that are typically extremely challenging to generate using quantum frequency processors \cite{lu2022high} (see Methods for the explicit expression of the learned projectors).

We truncate the Hilbert space dimension at the output of the QELM to form an $8\times8$ grid, in which the central $4\times4$ matrix initially encodes the amplitudes $|\alpha_{ij}|^2= |\alpha_i|^2\delta_{ij}$ of the input qudit states $\ket{ij}$.  
As for the qubit case, also here the QELM is trained using the classical process. 
Both the training and the test sets consist of balanced sets of qudit, qutrit and ququart states (see Methods for the detailed composition). 
Because of the large dimensionality of the QELM output, we choose to acquire also the test set via stimulated emission, performing frequency-resolved coincidence measurements only for the separable state $\ket{00}$ and for the three maximally entangled states $\ket{\Psi}=\frac{1}{\sqrt{M}}\sum_{j=1}^{M}\ket{jj}$ with $M=2,3,4$. 
Figure \ref{Fig_5}(a-c) shows examples of QELM outputs for various state dimensionalities and acquired through the stimulated process together with their expected outcomes, demonstrating excellent agreement, further quantified by an average fidelity of $\mathcal{F} = 0.997(2)$ between the full training and test sets and the corresponding simulations. 
The QELM outputs obtained from coincidence measurements exhibit a fidelity of $\mathcal{F} = 0.982(7)$ with the expected outcome. 

In Fig.~\ref{Fig_5}(d) we show the inferred values of the correlators $I_d$ across the test set, which align closely with their expected value ($\textrm{NMSE}=0.033(6)$). 
After classical training, the QELM is used to infer the SATWAP value for the spontaneously generated maximally entangled states, shown in Fig.~\ref{Fig_5}(e). 
Up to dimension three, the inferred $I_d$ exceeds the LHV bound by more than a standard deviation, while for dimension four the mean value remains above the classical bound. 
Overall, the classically trained QELM infers an increasing $I_d$ with the system dimensionality $d$. 
For all  states $\ket{\Psi}$, we find $I_d>Q_{d-1}$, where $Q_{d}=2(d-1)$ is the Tsirelson bound, the maximum value attainable by a $d$-dimensional quantum system \cite{wang2018multidimensional}. 
Consequently, the local dimension of each state is certified. 
\begin{figure*}[t!]    \includegraphics[width=\linewidth]{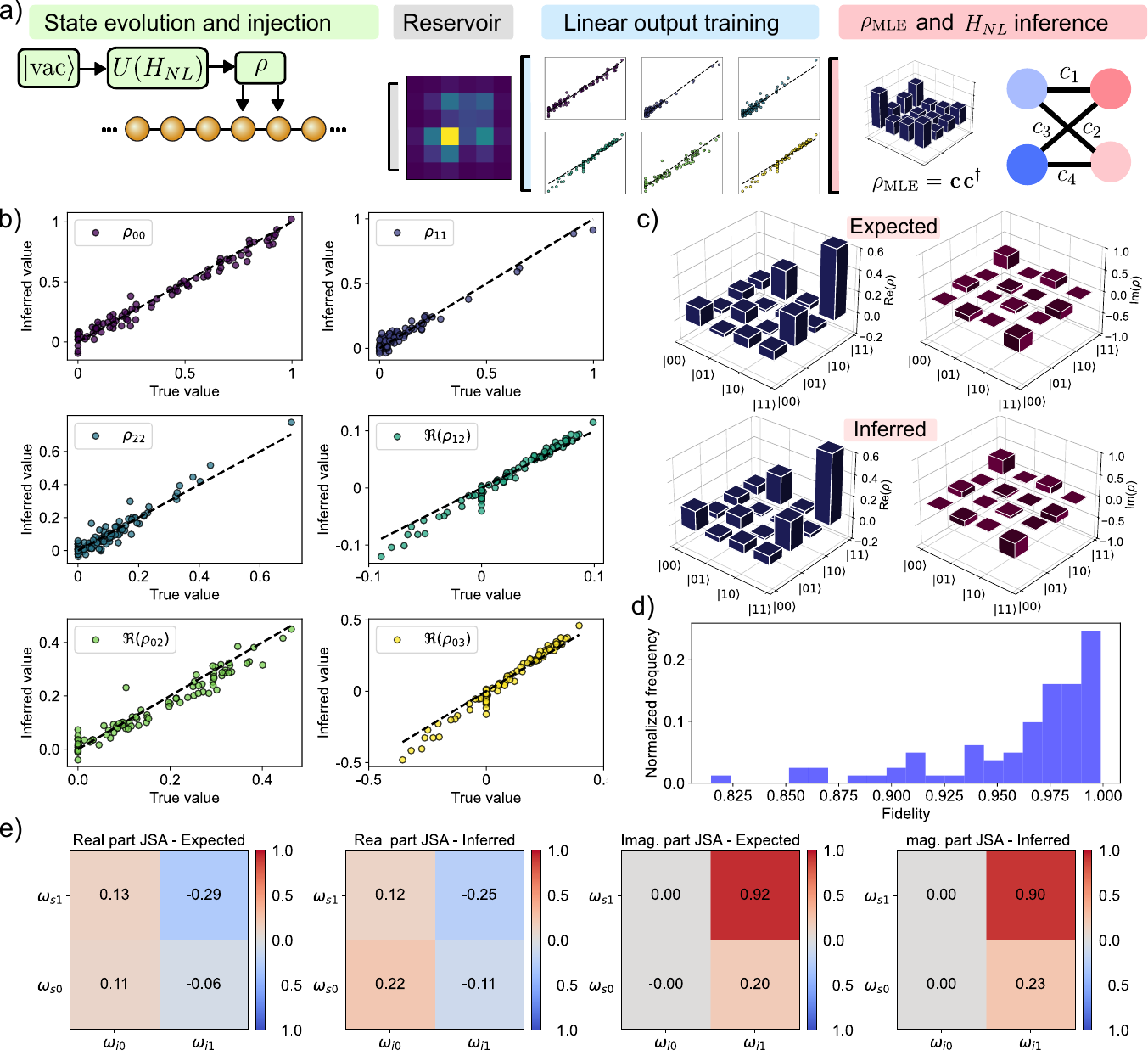}
    \caption{
    \textbf{Hamiltonian learning.}
    (a) Logical workflow for density matrix reconstruction and Hamiltonian learning using a QELM. The input state $\rho$ is prepared by evolving the input vacuum state through $H_{\textrm{NL}}$, and is fed to the reservoir, here represented by a node chain with nearest neighbour interaction. The state of each node is read and feeds the QELM, which is trained to reproduce six independent density matrix elements. The physical density matrix $\rho_{\textrm{MLE}}=\mathbf{c}\mathbf{c}^{\dagger}$ reproducing the elements is found by maximum likelihood, with $\mathbf{c}$ representing $H_{\textrm{NL}}$ on the computational basis in vectorized form. (b) Inferred versus true values of the six independent density matrix elements  characterizing the pure quantum states in the test set. (c) Example of inferred density matrix after maximum likelihood estimation (top panel) and its expected outcome (bottom panel). (d) Distribution of the fidelities between the inferred and the expected density matrices of the test set. (e) Example of reconstruction (Inferred) of  $H_{\textup{NL}}$ in the $2\times2$ signal-idler frequency-bin subspace and its expected outcome (Expected). 
    }
    \label{Fig_6}
\end{figure*}

\subsection*{\label{subsec:Nonlinear_Hamiltonian_learning}Hamiltonian learning}
\noindent
Having proved the inference capabilities of the system on nonclassical observables, we now benchmark the QELM in the task of learning the Hamiltonian responsible for the generation of photon-pairs, usually referred to as the nonlinear Hamiltonian of SpFWM.
In the low pump power regime, where the emission of multiple pairs is negligible, a close connection exists between the nonlinear Hamiltonian $\mathbf{H}_{\textup{NL}}=\sum_{qp} H_{qp}a^{\dagger}_q a^{\dagger}_p + \textrm{H.c.}$ and the state density matrix $\bm{\rho}$. 
Specifically, the state $\ket{\Psi}=e^{-\frac{i\mathbf{H}_{\textup{NL}}}{\hbar}}\ket{\textrm{vac}}$ is approximated to the first order in $\mathbf{H}_{\textup{NL}}$ as the biphoton state $\ket{\Psi}\sim \sum_{qp}H_{qp}\ket{qp}$, where the index $q(p)$ labels the $N$-dimensional computational basis of the signal(idler) photon. 
The density matrix $\bm{\rho}$, written on the basis $\ket{qp}$, is then $\bm{\rho}=\mathbf{c}\mathbf{c}^{\dagger}$, where $\mathbf{c}=\textrm{vec}(H)$. 
Note that the density matrix is parametrized by $2N^2-2$ real numbers because the state generated by $e^{-\frac{i\mathbf{H}_{\textup{NL}}}{\hbar}}$ is pure. 
As QELMs can be trained to infer the density matrix elements \cite{krisnanda2025experimental}, one can retrieve $\mathbf{c}$ (hence the elements $H_{qp}$) from the unique eigenvector of $\bm{\rho}$ with unit eigenvalue. 
The complete workflow for learning the nonlinear Hamiltonian is schematized in Fig.~\ref{Fig_6}(a). 

The input state, described by the target density matrix $\bm{\rho}_\mathrm{t}$, is coupled to the reservoir, producing its high-dimensional representation encoded in the output matrix of frequency correlations.
We include the waveshaper transformation $\mathbf{U}_{\textrm{ws}}$ into $\mathbf{H}_{\textup{NL}}$ to generate a variety of nonlinear Hamiltonians, i.e., $\mathbf{H}_{\textup{NL}}\mapsto\mathbf{U}_{\textrm{ws}}\mathbf{H}_{\textup{NL}}\mathbf{U}_{\textup{ws}}^{\dagger}$. 
The QELM is trained to infer the target $2N^2-2$ real parameters of the density matrix through multi-task linear regression. 
However, due to unavoidable noise and measurement errors, the reconstructed density matrix will likely be non-physical. 
Therefore, we look for the physical density matrix $\bm{\rho}_{\textup{MLE}}$ of rank-1 that has the maximum overlap with that learned by the QELM by using a standard maximum likelihood estimation technique (details are left to the Methods section). 
From the singular value decomposition of $\bm{\rho}_{\textup{MLE}}$, we calculate the unique eigenvector $\mathbf{c}$ and from this the elements of $\mathbf{H}_{\textup{NL}}$.

In the experiment, we implemented 512 training $\mathbf{H}_{\textup{NL}}$ by modifying the WS weights, and for each of them we acquired the frequency correlation matrix via stimulated emission. 
We focused on two-qubit states generated in the DP configuration. 
The number of independent parameters in the multi-task regression is then six. 
We arbitrarily choose $\mathbf{y}=\{\rho_{00},\rho_{11},\rho_{22},\mathcal{R}(\rho_{12}),\mathcal{R}(\rho_{02}),\mathcal{R}(\rho_{03}) \}$, where $\mathcal{R}(\cdot)$ denotes the real part. 
The test set is composed of $87$ states and has been acquired by coincidence measurements, following the same methodology described in Section \ref{subsec:QELM_training}. 
Figure \ref{Fig_6}(b) shows the inference on the test set after classical training of the QELM. The good inference performance is assessed from an average $\textrm{NMSE} =0.034$.

For each state, we then calculated the physical density matrix $\bm{\rho}_{\textup{MLE}}=\mathbf{c}\mathbf{c}^{\dagger}$ of rank-1 having the minimum MSE with the inferred set of elements in $\mathbf{y}$ (see Methods section for more details). An example of learned density matrix and its expected outcome is shown in Fig.~\ref{Fig_6}(c). 
As shown in Fig.~\ref{Fig_6}(d), the fidelity between the reconstructed density matrices and those expected from simulations is high and is on average $\mathcal{F}=0.96(4)$. 
By construction, this fidelity coincides with that between the expected and reconstructed matrices representing the nonlinear Hamiltonian in the computational basis. 
An example that showcases the good agreement between the QELM inference with the expected outcome is shown in Fig.~\ref{Fig_6}(e). 

As a conclusive remark, we point out that in the low-gain approximation where multiple-pair emission is neglected, $|H_{qp}|^2$ is the probability of detecting a signal photon in the frequency-bin $\ket{q}$ and the idler photon in the frequency-bin $\ket{p}$, thus it represents an element of the joint spectral intensity in the discrete basis set of the $N$ frequency-bins.
Therefore, the QELM correctly infers the phase-resolved joint spectral amplitude (i.e. the complex biphoton wavefunction) in the subset of frequency-bins describing the computational space of the signal-idler qubits.

\section*{\label{sec:Discussion}Discussion}
\noindent
\textcolor{black}{At the heart of the classical training procedure presented lies the ability to engineer a one-to-one correspondence between the transition probabilities of photon-pairs at the output of a linear optical network and the intensity distribution of classical fields at the output of the same network. When the two photons do not have spectral correlations in the frequency-bin basis, i.e. their joint state is separable, the correspondence can be trivially realized by using two independent coherent states. When the two photons do exhibit frequency-bin entanglement, spectral correlations are mimicked by structured twin beams generated by stimulated emission. Our experimental implementation encompasses both cases by cascading stimulated emission with a spectral waveshaper.} 
Due to these properties, the training weights can be transferred from the stimulated to the spontaneous regime. 
Remarkably, this strict correspondence enables not only qualitative inference about the state under test,  but also quantitative estimates, even involving nonclassical state properties such as its degree of entanglement. 
For example, we verified that the accuracy in inferring the value of the entanglement witness $\langle \mathcal{W} \rangle$ is comparable to that obtained from the reconstructed density matrix (see Supplementary Materials). 
Moreover, we confirmed that the QELM can learn the density matrix with fidelities relative to the expected states that can exceed those achieved via full quantum state tomography (see Supplementary Materials), notably using a single measurement setting instead of the nine local Pauli settings required in standard tomography.


\textcolor{black}{Our procedure is independent of the order of the spontaneous parametric process (second or third), number of bosonic modes, choice of degree of freedom, and pump temporal waveform.} 
Potential applications might include hyper-entanglement witnessing \cite{vallone2008hyperentanglement}, reconstruction of the joint spectral amplitude \cite{eckstein2014high}, learning of unitary maps \cite{krisnanda2025experimental}, of correlated quantum-walks \cite{imany2020probing} and reconstruction of density matrices \cite{rozema2015characterizing}.
While here we restricted our focus on the inference of linear functionals of the density matrix elements, such as the entanglement witness, nonlinear functionals of order $n$ could be in principle learned by the QELM by allowing $n$ copies of the input state to interact sequentially with the reservoir \cite{innocenti2023potential}. 
For unitary evolution of the joint input and reservoir state, this has been shown to be equivalent to performing specific POVMs on the input state $\rho^{\otimes n}$ \cite{innocenti2023potential}. 
In this perspective, we could prepare the signal and idler photons in the tensor-product state $\rho_\mathrm{s}\otimes\rho_\mathrm{i}$, with $\rho_\mathrm{s}=\rho_\mathrm{i}=\rho$, to estimate quadratic functionals of the density matrix, such as the state purity $\textrm{Tr}(\rho^2)$. 
Functionals of order $m_\mathrm{s}+m_\mathrm{i}$ could be, in principle, learned by encoding multiple qubits in each signal-idler photon through high-dimensional encoding \cite{baldazzi2025four}, such that $\rho_{\mathrm{s(i)}}=\rho^{\otimes m_{\mathrm{s(i)}}}$ for some integer $m_{\mathrm{s(i)}}>1$.\\
\textcolor{black}{Frequency-bin encoding is particularly appealing for quantum machine learning, combining the straightforward generation of frequency-entangled states \cite{kues2017chip} with efficient information scrambling within a single spatial mode.}
Indeed, combined with standard components such as waveshapers, EOMs, and tunable filters, frequency-bin-entangled photon sources enable the generation, evolution, and measurement of large classes of two-qubit states \cite{Lu2023}, here expressed (but in principle not limited to) by Eqs. (\ref{eq:SP_states}, \ref{eq:DP_states}).
All these components could furthermore be integrated on-chip, in the perspective of developing chip-scale quantum processing units \cite{clementi2023programmable, borghi2023reconfigurable, Nussbaum2022, Wu2025}.
Compared to other encoding schemes (e.g. path encoding), frequency-bins also provide robustness to mechanical drifts, which may critically affect long computations.
Moreover, they provide a natural way to scale the Hilbert space dimension to high-dimensional qudit states (here up to $d=4$ per photon), as well as a large reservoir dimension (here up to 8 per photon).
In this perspective, it is worth mentioning that the reservoir topology, here defined by a nearest-neighbor interaction, could be further refined to enhance the reservoir expressivity \cite{vetrano2025state}, for example via Fourier synthesis of complex driving RF tones, by the use of multiple EOMs, or by leveraging nonlinear conversion processes for frequency mixing.

\textcolor{black}{A natural question concerns the generalization of classical training to multiple qubits and qudits. At first sight, the connection between stimulated emission and photon-pair correlations might suggest that classical training is limited to two parties and to nonlinear optics in the low-gain regime. However, as we prove in the Supplementary Materials, classical training can be extended to an arbitrary number of parties and dimensions. This can be understood from two key observations. First, any operator $O$ acting on $N$ qubits can be expressed as a linear combination of the elements of an informationally complete POVM containing $4^N$ elements \cite{innocenti2023potential}. These elements can be chosen as tensor products of POVM elements generated by $N$ sub-reservoirs acting locally on each qubit. Second, owing to the linearity of the QELM approach, inference on previously unseen states is exact in principle when the training set consists of informationally complete quantum states and the reservoir is sufficiently expressive, as ensured by the construction described above.
Importantly, the informationally complete set of input states is not uniquely defined. For example, it can be constructed from tensor products of single-qubit states or from tensor products of single- and entangled-biphoton states. This latter scenario and the local sub-structure of the reservoir are directly relevant to the classical training scheme demonstrated here. In principle, training could therefore be performed using only independent coherent states, without relying on nonlinear optics, as partially explored in \cite{dibartolo2026efficientclassicaltrainingmodelfree}. The inclusion of correlated twin beams generated by stimulated emission, as demonstrated in our experiment, nevertheless provides greater flexibility in constructing the training set, with the potential to reduce the number of training states and improve the numerical stability of the resulting linear regression.
In addition, we show in the Supplementary Materials that classical training can be extended beyond the low-gain regime to the continuous-variable (CV) regime. We introduce a representative framework for the exact inference of CV observables that are at most quadratic in the quadratures of the input states. In this setting, the connection with stimulated emission is more fundamental. We show that stimulated emission provides nontrivial covariances and may improve the diversity and conditioning of the training set for the first- and second-order moments. Indeed, while all second-order moments of independent coherent states trivially factorize, nontrivial quadrature correlations emerge in the homodyne signals of conjugate twin beams.}\\ 

In conclusion, we have introduced and experimentally demonstrated a new paradigm for training QELMs which uses classical inputs to infer nonclassical features of unknown quantum states. 
The approach, here implemented on the frequency-bin basis, enables a drastic reduction in training time while simultaneously improving the signal-to-noise ratio, mitigating experimental drifts, and avoiding any resource overheads. 
We showcase excellent performance across multiple supervised-learning tasks, including entanglement witnessing, high-dimensional entanglement detection, and learning of the nonlinear Hamiltonian underlying  photon-pair generation.
\textcolor{black}{Looking ahead, we foresee extending this method beyond two qudits, with promising prospects for applications to continuous-variable Gaussian states and, more generally, for enabling faster and more robust training of machine-learning models for quantum feature extraction.} 

\subsection*{Funding and acknowledgments}
\noindent
D.B. acknowledges European Union funding from the STARLight project (project ID: 101194170). E.B., M.C., M.L., M.G. and M.B. acknowledge the PNRR MUR project PE0000023-NQSTI.
M.C. acknowledges funding from the European Union under the MSCA Postdoctoral Fellowship grant 101211100 (project GLINT).
The authors acknowledge CEA-Leti, Grenoble for providing the silicon photonic chip used as source of photon pairs.
The authors declare no competing interests.

\section*{METHODS}
\subsection*{Experimental setup}
\noindent
The pump is a tunable continuous-wave laser that is passed through an EOM driven by a sinusoidal signal at $20$ GHz with an RF power of $23$ dBm. Phase modulation generates multiple electro-optic comb lines. A WS is then used to select either one ($\lambda_{\mathrm{p}}\sim1556.56$ nm) or two ($\lambda_{\mathrm{p}_{\mathrm{1}}} = 1556.48$ nm and $\lambda_{\mathrm{p}_{\mathrm{2}}} = 1556.64$ nm) comb lines for the generation of SP and DP states, respectively. The pump is subsequently amplified using an Erbium-Doped Fiber Amplifier, and its polarization is set to TE. In the DP configuration, the powers of the two selected comb lines are adjusted such that the ratio between the non-degenerate and degenerate SpFWM processes is approximately $4$. A bandpass filter is then used to remove background noise at the wavelengths of the signal and idler photons.
During the training phase, a tunable seed laser is employed to stimulate the FWM process. This laser is phase-modulated at $20$ GHz with an RF power of $23$ dBm using an EOM, and subsequently passed through a WS that applies the idler weights $\mathbf{g}_\mathrm{i}$. The polarization of the seed laser is aligned with that of the pump, and the two beams are multiplexed into a common channel of a fiber array. Light is coupled into and out of the silicon photonic chip (fabricated by CEA-Leti Grenoble) using a pair of grating couplers, resulting in a total insertion loss of approximately $\ 6$ dB and an effective on-chip pump and seed powers of $\sim15$ dBm and $\sim5$ dBm respectively. \\
The photon-pair source consists of a rib waveguide with a length of $1.6$ cm, a width of $450$ nm, a total height of $310$ nm, and a shallow etch depth of $155$ nm. The source brightness is measured to be $3\cdot10^6\,\textrm{Hz}/(\textrm{mW}^2\cdot\textrm{nm})$.\\
In the spontaneous process, discrete frequency-bins are carved from the broad-band photon-pair emission using the WS, each with a width and spacing of $20$ GHz. The signal and idler bins are centered at $\lambda_{\mathrm{s},n}=1561.43+n\Delta$ and $\lambda_{\mathrm{i},n} = 1551.72-n\Delta$, respectively, with $n\in Z$ and $\Delta=0.16$ nm. The WS is used both to carve the frequency-bins and to apply the complex weights $\mathbf{g}$. Qubit states are prepared by allowing only two frequency-bins to pass while blocking all other wavelengths, whereas up to four bins are transmitted to generate qudit states.
In the DP configuration, the state of the signal-idler photons at the output of the WS is given by Eq.(\ref{eq:DP_states}). In the SP configuration, the resulting states take the form:
\begin{equation}
    \ket{\Psi}_{\textup{SP}}=
    \frac{1}{\mathcal{N}_{\textup{SP}}(\mathbf{g})}
    \left( g_{\mathrm{s_0}}g_{\mathrm{i_0}}\ket{00}+g_{\mathrm{s_1}}g_{\mathrm{i_1}}\ket{11}
    \right).
    \label{eq:SP_states}
\end{equation}
Note that Eq.~\eqref{eq:SP_states} and Eq.~\eqref{eq:DP_states} cannot be made equivalent for any choice of the coefficients $\mathbf{g}$; therefore, they span distinct subsets of the two-qubit Hilbert space. 
While the coefficients of the four logical states are not all independent, the family of states in Eq.(\ref{eq:DP_states}) span a large portion of two-qubit Hilbert space, which is sufficient for demonstrating the generalization capabilities of the QELM discussed in the main text.\\ 
After the WS, the signal and idler photons exit through two separate fibers, which are directed to independent EOMs driven by sinusoidal waveforms at $20$ GHz with an RF power of $23$ dBm. In the stimulated process, only the signal beam is passed through the WS and the EOM. A passband filter is used in both the signal and idler paths to suppress residual pump light.
The correlation matrices shown, for example, in Fig.~\ref{Fig_3}(b,c) are obtained in the spontaneous process via frequency-resolved coincidence detection. Fiber Bragg gratings with a bandwidth of $12$ GHz, together with circulators, are used to select the desired frequency-bin combinations, which are detected by pairs of superconducting nanowire single-photon detectors with a detection efficiency of $85\%$. In the stimulated process, the spectra of the signal beam are recorded using an Optical Spectrum Analyzer.

\subsection*{QELM training}\label{sec:QELM_training}
\noindent
We detail here the methods used to train the QELM for reproducing the results presented in the main manuscript. For each input state $r=(1,...,N_{\mathrm{in}})$, the corresponding output from the QELM is a $Q\times Q$ frequency-resolved correlation matrix $\{C_{kj}\}_{k,j=0}^{Q-1}$, which is reshaped into a column vector $[\mathbf{x}^{(r)}]^T$ of length $Q^2$. Each vector is rescaled to have unit norm and then placed into the $r^{\textup{th}}$ column of the $Q^2\times N_{\mathrm{in}}$ matrix $\mathbf{X}$. Training consists of determining the $M\times Q^2$ weight matrix $\mathbf{W}$ that minimizes the mean squared error $||\mathbf{Y}-\mathbf{W}\mathbf{X}||^2$, where $\mathbf{Y}$ is an $M\times N_{\mathrm{in}}$ matrix of $M$ targets labels.
\\
\emph{Entanglement witness} \\
Training is performed by regularized Elastic Net linear regression \cite{zou2005regularization}. The $\ell_1$ regularization strength is selected from the grid  $[0.1,0.5,0.7]$.
The regularization parameter is chosen from a list of 100 logarithmically spaced values between $10^{-5}$ and $10$, and is optimized through 9-fold cross validation using the $R^2$ score as the evaluation metric. Before training, both the training and test datasets are pre-processed by standardizing the input feature matrix. The training dataset consists of 290 states, composed of 70 separable states, 70 states from the SP configuration, and 150 states from the DP configuration. The test set consists of 96 states, including 16 separable states, 35 states from the SP configuration, and 45 states from the DP configuration. Both the training and test sets are randomly sampled from a pool of 250 states prepared in the SP configuration and 500 states prepared in the DP configuration. \\
\emph{SATWAP inequalities} \\
Definition: the SATWAP inequality can be defined on two systems $A$ and $B$ sharing a quantum state. On their state, each party performs one of two measurements $A_1,A_2$ and $B_1,B_2$, each yielding one of $d$ possible outcomes labelled by $0,...,d-1$. The inequality is written as \cite{wang2018multidimensional}
\begin{equation}
\begin{aligned}
I_d := &\sum_{l=1}^{d-1} \;
a_l\langle{A_1^lB_1^{d-l}}\rangle
+ a_l^*\omega^l\langle{A_1^lB_2^{d-l}} \rangle\\
&+ a_l\langle{A_2^lB_2^{d-l}}\rangle
+ a_l^*\langle{A_2^lB_1^{d-l}}\rangle
\le C_d,
\end{aligned}
\label{eq:SATWAP}
\end{equation}
where $\omega =e^{\frac{2\pi i}{d}}$, $a_l=\omega^{(2l-d)/8}/\sqrt{2}$ and $C_d$ is its classical bound given by:
\begin{equation}
    C_d = \frac{1}{2} \left [3 \cot \left (\frac{\pi}{4d} \right) - \cot \left (\frac{3\pi}{4d} \right)\right ]-2. \label{eq:LHV_bound}
\end{equation}
The observables $A_{1(2)}$ and $B_{1(2)}$ have eigenprojectors 
\begin{equation}
\begin{aligned}
\ket{a}_{1(2)} 
&= \frac{1}{\sqrt{d}}\sum_{k=0}^{d-1}
e^{\frac{i2\pi k(a-\theta_{1(2)})}{d}}\ket{k}, \\
\ket{b}_{1(2)} 
&= \frac{1}{\sqrt{d}}\sum_{k=0}^{d-1}
e^{\frac{i2\pi k(\xi_{1(2)}-b)}{d}}\ket{k},
\end{aligned}
\end{equation}
where $a,b\in\{0,...,d-1\}$, $\theta_1=1/4$, $\theta_2 = 3/4$, $\xi_1 = 1/2$ and $\xi_2 = 1$. \\ 
Training procedure: training is performed using regularized Elastic Net linear regression. The $\ell_1$ regularization strength is selected from the grid $[0,0.05,0.1,0.5,0.7,1]$.
The regularization strength is chosen from 100 logarithmically spaced values ranging from $10^{-9}$ to $10^{-3}$, and is optimized by 4-fold cross-validation using the $R^2$ as a selection metric. Prior to training, both the training and test datasets are pre-processed through feature selection using a variance-threshold criterion set to $5\cdot 10^{-4}$. This reduces the dimensionality of the input from $64$ to $21$ features, which are then standardized before model fitting. The training set consists of $216$ states, divided into four subsets. In the $k^{\textup{th}}$ subset, $k-1$ of the coefficients $\alpha_j$ of the input state $\ket{\Psi}$ (see main text) are set to zero, while the nonzero coefficients are assigned randomly as $\alpha_j=|\alpha_j|e^{i\phi_j}$ with $|\alpha_j|\sim\mathcal{U}[0,1]$ and $\phi_j\sim\mathcal{U}[-\pi,\pi]$. 
The test set consists of four subsets of $26$ states each. This stratification ensures that neither the training set nor the test set is biased toward any particular qudit dimension.\\
\emph{Density matrix reconstruction} \\
Training is performed by multitask regularized Elastic Net linear regression \cite{pedregosa2011scikit}. The $\ell_1$ regularization strength is selected from the grid  $[0.01,0.05,0.1, 0.25,0.5, 0.9,1]$.
The overall regularization parameter is chosen from 50 logarithmically spaced values from $10^{-6}$ to $1$, and is optimized by 5-fold cross validation using $R^2$ score as the metric.
Prior to training, the input data is pre-processed by feature selection using a variance threshold criterion set to $2.5\cdot 10^{-5}$. This reduces the number of input features from $36$ to $21$. The selected features are subsequently standardized.\\
The QELM is trained to infer six independent elements of the density matrix for each input state, $\mathbf{y}=\{\rho_{00},\rho_{11},\rho_{22},\mathcal{R}(\rho_{12}),\mathcal{R}(\rho_{02}),\mathcal{R}(\rho_{03}) \}$. To obtain the physical rank-1 density matrix $\bm{\rho}_{\textrm{MLE}}$ that most likely reproduces the inferred elements in $\mathbf{y}$, we first parametrize the four-component vector $\mathbf{c}$ using six real parameters $\mathbf{q}$:
\begin{equation}
    \begin{split}
        c_1 & = q_0 \\
        c_2 & = \sqrt{1-q_0^2} e^{iq_1}\cos(q_2) \\
        c_3 & = \sqrt{1-q_0^2}e^{i(q_1+q_3)}\sin(q_2)\cos(q_4) \\
        c_4 & = \sqrt{1-q_0^2}e^{i(q_4+q_5)}\sin(q_2)\sin(q_4)
    \end{split} 
    \label{eq:theta_parametrization}
\end{equation}
with $q_0\in[0,1]$, $q_2,q_4\in [0,\frac{\pi}{2}]$ and $q_1,q_3,q_5 \in [0,2\pi)$. This parametrization ensures that $||\mathbf{c}||^2=1$. The corresponding physical density matrix is constructed as   $\bm{\rho}_{\textrm{MLE}}=\mathbf{c}\mathbf{c}^{\dagger}$, which is, by construction, rank-1 and therefore represents a pure quantum state. Finally, the parameters $\mathbf{q}$ are obtained by minimizing the maximum-likelihood cost function $||\mathbf{y}-\tilde{\mathbf{y}}(\mathbf{q})||^2$, using Particle Swarm optimization, where $\tilde{\mathbf{y}}$ denotes the corresponding density matrix elements extracted from $\bm{\rho}_{\textup{MLE}}$.

\setcounter{figure}{0}
\renewcommand{\thefigure}{S\arabic{figure}}
\setcounter{equation}{0}
\renewcommand{\theequation}{S\arabic{equation}}
\renewcommand{\theHfigure}{S\arabic{figure}}
\renewcommand{\theHequation}{S\arabic{equation}}
\section*{SUPPLEMENTARY MATERIALS}
\textcolor{black}{
\subsection*{General conditions for QELM training with classical inputs and discrete variables}
\noindent
In this section, we show that a QELM acting on $N$ qubits can be trained using only classical inputs that are evolved under $N$ separable reservoir systems. We first derive the general conditions on the $Q$ effective POVMs $\{\mu_j\}_{j=1}^{Q}$ and the $R_{\textrm{tr}}$ training states $\{\rho^{\textrm{tr}}_{j}\}_{j=1}^{R_{\textrm{tr}}}$ that allow for the unambiguous reconstruction of the expectation values of a generic $D=4^N$-dimensional Hermitian operator $O$ acting on the joint state of $N$ qubits. Then, we demonstrate that these conditions can be satisfied by the classical training method presented in the main text.\\
The training stage consists of finding the solution to the linear system
\begin{equation}
\mathbf{y}_{\textrm{tr}}=\langle \bm{\mu},\bm{\rho}_{\textrm{tr}}\rangle^{T}\mathbf{w}, \label{eq:training_equation}    
\end{equation}
where $\mathbf{y}_{\textrm{tr}}=\langle O,\bm{\rho}^{\textrm{tr}}\rangle^{T}$ is the length-$R_{\textrm{tr}}$ column vector of labels, $\langle A,B\rangle=\textrm{Tr}(A^{\dagger}B)$ denotes the Hilbert-Schmidt inner product, and $\mathbf{w}$ is the length-$Q$ vector of unknown weights. The matrix entries of $\langle \bm{\mu},\bm{\rho}_{\textrm{tr}}\rangle$ are given by $[\langle \bm{\mu},\bm{\rho}_{\textrm{tr}}\rangle]_{mn}=\textrm{Tr}(\mu_m^{\dagger}\rho^{\textrm{tr}}_n)=\textrm{Tr}(\mu_m\rho^{\textrm{tr}}_n)$.
Typically, $R_{\textrm{tr}}>Q$, which implies that $\langle \bm{\mu},\bm{\rho}_{\textrm{tr}}\rangle^{T}$ is an $R_{\textrm{tr}}\times Q$ tall matrix. The solution in Eq.~(\ref{eq:solution_via_pseudo_inverse}) exists provided that $\mathbf{y}_{\textrm{tr}}\in\textrm{Im}(\langle \bm{\mu},\bm{\rho}_{\textrm{tr}}\rangle^{T})$, where $\textrm{Im}(\cdot)$ denotes the image (column) space. Moreover, the solution is unique if $\textrm{rk}(\langle \bm{\mu},\bm{\rho}_{\textrm{tr}}\rangle^{T})=\textrm{min}(R_{\textrm{tr}},Q)=Q$, i.e., if $\langle \bm{\mu},\bm{\rho}_{\textrm{tr}}\rangle^{T}$ has full rank, whereas infinitely many solutions exist if $\langle \bm{\mu},\bm{\rho}_{\textrm{tr}}\rangle^{T}$ is rank-deficient. We can equivalently express the condition $\mathbf{y}_{\textrm{tr}}\in\textrm{Im}(\langle \bm{\mu},\bm{\rho}_{\textrm{tr}}\rangle^{T})$ as 
\begin{equation}
    \textrm{Tr}(O\rho_i^{\textrm{tr}})=\textrm{Tr}\left [\left (\sum_{q=1}^{Q}c_q\mu_q \right)\rho^{\textrm{tr}}_i\right ]\qquad i=1,..,R_{\textrm{tr}}, \label{eq:trace_equality}
\end{equation}
for some coefficients $c_q$. By defining 
\begin{equation}
    A = O-\sum_{q=1}^{Q}c_q\mu_q, \label{eq:A_operator}
\end{equation}
we can rewrite Eq.~(\ref{eq:trace_equality}) as 
\begin{equation}
    \textrm{Tr}(A\rho_{i}^{\textrm{tr}})=0 \qquad \forall i. \label{eq:A_condition}
\end{equation}
If the set of training states $\{\rho_{i}^{\textrm{tr}}\}_{i=1}^{R_{\textrm{tr}}}$ is informationally complete, then Eq.~(\ref{eq:A_condition}) necessarily implies \mbox{$O=\sum_{q=1}^{Q}c_q\mu_q$} (and vice versa), i.e., $O\in \textrm{span}(\{\mu_j\}_{j=1}^{Q})$. Otherwise, we can only conclude that $O$ and $\sum_{q=1}^{Q}c_q\mu_q$ act identically on the subspace $\textrm{span}(\{ \rho_{j}^{\textrm{tr}}\}_{j=1}^{R_{\textrm{tr}}})$ spanned by the training states. 
The minimum-norm solution to Eq.~(\ref{eq:training_equation}) is expressed through the pseudoinverse $(\langle \bm{\mu},\bm{\rho}_{\textrm{tr}}\rangle^{T})^{+}=(\langle \bm{\mu},\bm{\rho}_{\textrm{tr}}\rangle^{+})^{T}$ as 
\begin{equation}
    \mathbf{w} = (\langle \bm{\mu},\bm{\rho}_{\textrm{tr}}\rangle^{T})^{+}\mathbf{y}_{\textrm{tr}}. \label{eq:solution_via_pseudo_inverse}
\end{equation}
Having assessed the conditions and the related physical implications for the existence of the pseudoinverse, we now evaluate the error in the inferred value of $O$ for a new trial state $\rho^{\textrm{new}}$. In general, this state can be written as $\rho^{\textrm{new}}=\rho_{\parallel}+\rho_{\perp}$, where $\rho_{\parallel}\in\textrm{span}(\{\rho_{j}^{\textrm{tr}}\}_{j=1}^{R_{\textrm{tr}}})$ and $\rho_{\perp}$ belongs to its orthogonal complement. By definition, we can write $\rho_{\parallel}=\sum_{j=1}^{R_{\textrm{tr}}}\eta_j^{\textrm{tr}}\rho_{j}^{\textrm{tr}}$, where $\{\eta_j\}_{j=1}^{R_{\textrm{tr}}}$ is a set of expansion coefficients. Similarly, we can expand $\rho_{\perp}$ in terms of the $R_{\perp}$ basis vectors $\{\rho^{\perp}_j\}_{j=1}^{R_{\perp}}$ spanning the orthogonal complement as $\rho_{\perp}=\sum_{j=1}^{R^{\perp}}\eta_j^{\perp}\rho_{j}^{\perp}$. 
The value of $O$ inferred by the QELM for the new state is $y_{\textrm{new}}=(\langle \bm{\mu},\rho^{\textrm{new}}\rangle)^T\mathbf{w}$, and by combining this with Eq.~(\ref{eq:solution_via_pseudo_inverse}), we get
\begin{equation}
    y_{\textrm{new}}=(\bm{\eta}^{\textrm{tr}})^T\langle \bm{\mu},\bm{\rho}_{\textrm{tr}}\rangle^T(\langle \bm{\mu},\bm{\rho}_{\textrm{tr}}\rangle^T)^{+}\mathbf{y}_{\textrm{tr}}+\langle \bm{\mu},\rho_{\perp}\rangle^T\mathbf{w}.
\end{equation}
Since, by hypothesis, $\mathbf{y}_{\textrm{tr}}\in\textrm{Im}(\langle \bm{\mu},\bm{\rho}_{\textrm{tr}}\rangle^{T})$, and $\langle \bm{\mu},\bm{\rho}_{\textrm{tr}}\rangle^T(\langle \bm{\mu},\bm{\rho}_{\textrm{tr}}\rangle^T)^{+}$ is an orthogonal projector onto that subspace, the first term on the right-hand side simply reduces to $(\bm{\eta}^{\textrm{tr}})^T\mathbf{y}_{\textrm{tr}}$. Therefore, we have
\begin{equation}
    y_{\textrm{new}}=((\bm{\eta}^{\textrm{tr}})^T+\langle\bm{\mu},\rho_{\perp}\rangle^T(\langle \bm{\mu},\bm{\rho}_{\textrm{tr}}\rangle^T)^{+})\mathbf{y}_{\textrm{tr}}. \label{eq:y_new_expanded}
\end{equation}
The exact expectation value $y_{\textrm{true}}=\langle O,\rho^{\textrm{new}}\rangle$ can be written using the state expansion above as 
\begin{equation}
    y_{\textrm{true}} = (\bm{\eta}^{\textrm{tr}})^T\mathbf{y}_{\textrm{tr}}+\langle O,\rho_{\perp}\rangle. \label{eq:y_true}
\end{equation}
We can subtract Eq.~(\ref{eq:y_new_expanded}) from Eq.~(\ref{eq:y_true}) to obtain the error $\Delta y=y_{\textrm{true}}-y_{\textrm{new}}$ in the inference stage as
\begin{equation}
    \Delta y = \langle O,\rho_{\perp}\rangle-\langle\bm{\mu},\rho_{\perp}\rangle^T(\langle \bm{\mu},\bm{\rho}_{\textrm{tr}}\rangle^T)^{+}\mathbf{y}_{\textrm{tr}} \label{eq:residual_error}
\end{equation}
}
\begin{figure*}[t!]    \includegraphics[width=0.75\textwidth]{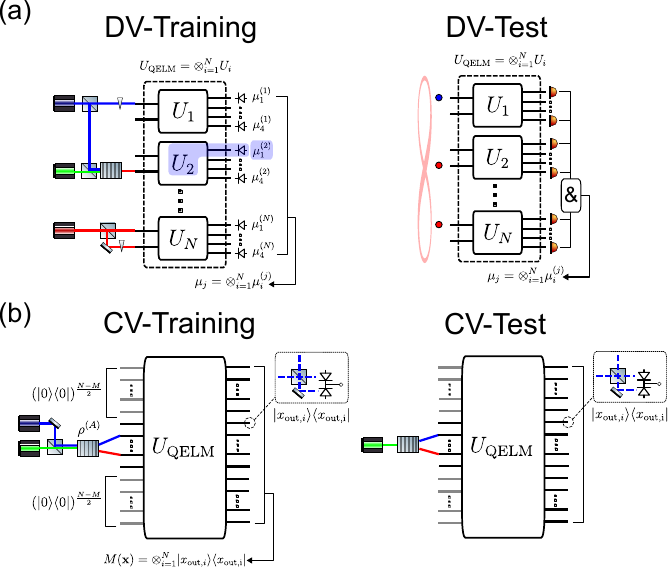}
    \caption{\textcolor{black}{(a) Left: Classical training of a discrete-variable QELM acting on $N$ qubits. Input states can be either independent coherent states (red), mimicking the output statistics of single photons, or twin beams (blue and red) produced by stimulated emission from a nonlinear crystal pumped by a strong beam (green), mimicking the output statistics of entangled photon pairs. The reservoir is divided into $N$ independent sub-reservoirs described by local unitary evolutions $U_{j=1,\dots,N}$. Here, the number of output modes for each sub-reservoir is set to $4$. The intensity of the output beams is recorded by photodiodes. The QELM evolution, followed by intensity detection on the output modes, corresponds to an effective POVM $\mu$ performed on the qubit at the input of the reservoir. Right: Inference of quantum states on the QELM previously trained with classical inputs. The input is a generic $N$-qubit state encoded using $N$ single photons. The output modes are monitored by $N$-fold coincidence detection using an array of single-photon detectors. (b) Left: Classical training of a continuous-variable QELM. The input state is encoded in $M$ bosonic modes, which evolve through a generic Gaussian transformation defined by the unitary operator $U_{\textrm{QELM}}$. Ancillary $N-M$ modes, initialized in the vacuum state, are introduced at the input. The $N$ output modes are measured via homodyne detection. The training states are generally multimode states given by tensor products of coherent states and twin beams generated by seeded parametric amplification. Right:  Inference of CV states on the QELM previously trained with classical inputs. These may include squeezed-vacuum states and non-Gaussian states.} 
    }
    \label{Fig:S1}
\end{figure*}
\textcolor{black}{
To provide further insight into the error term in Eq.~(\ref{eq:residual_error}), we represent each term using the vectorization of the operators in a $D$-dimensional orthonormal basis under the Hilbert-Schmidt inner product. The first term on the right-hand side of Eq.~(\ref{eq:residual_error}) becomes \mbox{$\langle O,\rho_{\perp} \rangle=\textrm{vec}(\mathbf{O})^T \mathbf{G}_{\perp}\bm{\eta}_{\perp}$}, where $\mathbf{G}_{\perp}$ is a $D\times R_{\perp}$ matrix obtained by stacking $\{\textrm{vec}(\rho^{\perp}_j)\}_{j=1}^{R_\perp}$ column-wise. Similarly, the second term becomes $(\mathbf{MG}_{\perp}\bm{\eta}_{\perp})^T((\mathbf{MG}_{\parallel})^T)^{+}\mathbf{y}_{\textrm{tr}}$, where $\mathbf{G}_{\parallel}$ is a $D\times R_{\textrm{tr}}$ matrix obtained by stacking $\{\textrm{vec}(\rho_{j}^{\textrm{tr}})\}_{j=1}^{R_{\textrm{tr}}}$ column-wise, and $\mathbf{M}$ is a $Q\times D$ matrix obtained by stacking $\{\textrm{vec}(\mu_j)^T\}_{j=1}^{Q}$ row-wise. By grouping the common factor $\bm{\eta}_{\perp}$ we then obtain
\begin{equation}
    \Delta y = \bm{\eta}_{\perp}^T\mathbf{G}_{\perp}^T(\textrm{vec}(\mathbf{O})-\mathbf{M}^T((\mathbf{M} \mathbf{G}_{\parallel})^T)^{+}\mathbf{y}_{\textrm{tr}}).
\end{equation}
By expressing $\mathbf{y}_{\textrm{tr}}=\mathbf{G}_{\parallel}^T\textrm{vec}(\mathbf{O})$ and grouping terms, we obtain the compact final form
\begin{equation}
    \Delta y = \bm{\eta}_{\perp}^T\mathbf{G}_{\perp}^T(\mathbf{I}-(\mathbf{G}_{\parallel}(\mathbf{MG}_{\parallel})^+\mathbf{M})^T)\textrm{vec}(\mathbf{O}), \label{eq:final_error_expression}
\end{equation}
where $\mathbf{I}$ denotes the identity matrix. 
If $\{\rho^{\textrm{tr}}_j\}_{j=1}^{R_{\textrm{tr}}}$ forms an informationally complete set or \mbox{$\rho^{\textrm{new}}\in\textrm{span}(\{\rho_j^{\textrm{tr}}\}_{j=1}^{R_{\textrm{tr}}})$}, then  $\bm{\eta}_{\perp}=0$ and $\Delta y=0$. On the other hand, if $\rho^{\textrm{new}}\notin \textrm{span}(\{\rho_j^{\textrm{tr}}\}_{j=1}^{R_{\textrm{tr}}})$ because $\textrm{dim}(\textrm{span}(\{\rho_j^{\textrm{tr}}\}_{j=1}^{R_{\textrm{tr}}}))<D$ (or equivalently $\textrm{rk}(\mathbf{G}_{\parallel})<D$), then the condition $\Delta y=0$, for a general $O$, requires that the term in brackets in Eq.~(\ref{eq:final_error_expression}) vanishes. However, since $\textrm{rk}(\mathbf{G}_{\parallel}(\mathbf{MG}_{\parallel})^+\mathbf{M})\leq\textrm{rk}(\mathbf{G}_{\parallel})<D$ by hypothesis, and $\textrm{rk}(\mathbf{I})=D$, this condition is not in general fulfilled.\\
In summary, a QELM implementing a set of effective POVMs $\{\mu_j\}_{j=1}^{Q}$ and trained on the set of states $\{\rho^{\textrm{tr}}_j\}_{j=1}^{R_{\textrm{tr}}}$ with labels $\mathbf{y}_{\textrm{tr}}=\langle O,\bm{\rho}_{\textrm{tr}} \rangle^T$ can exactly infer the expectation value of an observable $O$ for a new test state $\rho^{\textrm{new}}$ if $\mathbf{y}_{\textrm{tr}}\in \textrm{Im}(\langle \bm{\mu},\bm{\rho}_{\textrm{tr}}\rangle^T)$ and $\rho^{\textrm{new}}\in\textrm{span}(\{\rho_j^{\textrm{tr}}\}_{j=1}^{R_{\textrm{tr}}})$. Additionally, a specific $O$ can be reconstructed if $O\in\textrm{span}(\{\mu_j\}_{j=1}^Q)$, which implies that if the set of POVMs $\{\mu_j\}_{j=1}^Q$ is informationally complete, any $D$-dimensional Hermitian operator can be reconstructed. \\
The most relevant observation is that the set $\{\rho^{\textrm{tr}}_j\}_{j=1}^{R_{\textrm{tr}}}$ of training density matrices can consist entirely of separable states of the form $\rho_j^{\textrm{tr}}=\bigotimes_{i=1}^{N}\rho_i^{(j)}$. Moreover, the POVMs themselves can be constructed from $N$ independent POVMs acting locally on the $N$ qubits, i.e., $\mu_j=\bigotimes_{i=1}^{N}\mu^{(j)}_i$. For example, one can engineer $N$ local reservoirs, each implementing four POVM elements $\{\mu_i^{(j)}\}_{i=1}^{4}$ on each qubit space which form an informationally complete set, yielding $\textrm{dim}(\textrm{span}(\{\mu_j\}_{j=1}^{D}))=4^N$. A scheme of this architecture and of the workflow from classical training to the inference on new genuine quantum states is sketched in Fig. \ref{Fig:S1}(a). Since the outcome probabilities of separable states evolved through local reservoirs can be evaluated using only classical intensity measurements on coherent states \cite{laing2012super}, we conclude that the classical training of a photonic QELM, presented for two qubits in the main manuscript, can indeed be extended to an arbitrary number of qubits. Remarkably, since the above derivation does not depend on the local dimension of each photon, the approach straightforwardly extends to $N$ qudits.\\
While the matrix $\langle \bm{\mu},\bm{\rho}_{\textrm{tr}}\rangle$ can, in principle, be evaluated using only separable states at the input, the use of stimulated emission offers two key advantages
\begin{enumerate}
    \item \emph{Reduced number of training states.} Exact inference on arbitrary new test states requires the training set to span the relevant state space, whose dimension remains $4^N$. However, the structure of the training set determines how efficiently specific classes of test states can be represented within this space. When the new inputs exhibit strong nonclassical correlations, a training set composed solely of separable states may require a large number of states to efficiently capture these correlations. By including bipartite nonclassical correlations, such as those mimicked by stimulated emission, the training set explores additional directions within the same $4^N$-dimensional space. This can provide a more efficient representation of correlated test states and, for specific classes of entangled inputs, potentially reduce the number of training states required, without reducing the dimensionality of the underlying state space.
    \item \emph{Improved numerical stability.} The weight vector $\mathbf{w}$ is calculated by solving the linear system in Eq.~(\ref{eq:training_equation}) via the pseudoinverse (Eq.~(\ref{eq:solution_via_pseudo_inverse})); hence, the numerical stability of the solution is determined by the condition number $\kappa(\langle \bm{\mu},\bm{\rho}_{\textrm{tr}}\rangle)=\left | \frac{\sigma_{\textrm{max}}}{\sigma_{\textrm{min}}}\right |$, where $\sigma_{\textrm{max}}$ and $\sigma_{\textrm{min}}$ denote the maximum and minimum singular values of the matrix $\langle \bm{\mu},\bm{\rho}_{\textrm{tr}}\rangle$. In particular, the relative error $\Delta\mathbf{w}/\mathbf{w}$ in the training coefficients is bounded above by \cite{innocenti2023potential}
    \begin{equation}
        \frac{||\Delta \mathbf{w}||}{||\mathbf{w}||} \leq\kappa(\langle \bm{\mu},\bm{\rho}_{\textrm{tr}}\rangle)  \frac{||\Delta\mathbf{y}_{\textrm{tr}}||}{||\mathbf{y}_{\textrm{tr}}||} . \label{eq:condition_number_bound}
    \end{equation}
    For a fixed reservoir configuration characterized by the set of POVMs $\bm{\mu}$, including bipartite correlations in the training set $\bm{\rho}_{\textrm{tr}}$ introduces greater flexibility and diversity into the entries of the matrix $\langle \bm{\mu},\bm{\rho}_{\textrm{tr}}\rangle$. This can be leveraged to decrease the condition number, thereby improving the numerical stability of the solution. This is particularly relevant for reservoirs where the local POVMs $\bm{\mu}$ yield very similar outcomes for different input training states, introducing small singular values that do not convey physically meaningful information.\\
    Along similar lines, when training using only separable states, their mathematical overlap $\langle O,\bm{\rho}_{\textrm{tr}} \rangle $ with highly non-local observables $O$ may be very small, thereby reducing the signal-to-noise ratio $|\mathbf{y}_{\textrm{tr}}/\Delta\mathbf{y}_{\textrm{tr}}|$. Note that even if the target labels $\mathbf{y}_{\textrm{tr}}$ are not measured experimentally, they inherit errors arising from state-preparation imperfections. For instance, if an imperfect copy $\tilde{\rho}^{\textrm{tr}}=\rho^{\textrm{tr}}+\delta\rho$ of the target state $\rho^{\textrm{tr}}$ is prepared, the error in the assigned label will be $\Delta y_{\textrm{tr}}=\langle O,\delta\rho\rangle$. 
\end{enumerate}
}
\textcolor{black}{
\subsection*{Classical training of a continuous-variable QELM}
\noindent
The aim of this section is to show that a QELM trained using a combination of coherent states and correlated twin beams generated by stimulated emission can infer properties of previously unseen continuous-variable (CV) nonclassical states, such as squeezed-vacuum states and other non-Gaussian states. This generalizes the approach presented in the main text, which relies on a discrete-variable description, to the CV domain. We focus on a specific class of tasks: inferring expectation values of observables $\langle O \rangle$ that are at most quadratic in the quadrature operators, while restricting our attention to Gaussian training states. This simplified framework enables a clear analytical treatment of the problem and allows us to identify the algebraic conditions under which $\langle O \rangle$ can be retrieved exactly. At the same time, it remains applicable to a broad class of observables which are relevant in CV quantum optics. \\
We consider a system described by a density matrix  $\rho\in\mathcal{H}_A\otimes \mathcal{H}_V$ composed of $N$ bosonic modes, where $M<N$ modes host the state of interest $\rho^{(A)}\in \mathcal{H}_A$, and $N-M$ ancillary reservoir modes are initialized in the vacuum state, i.e., $\rho=\rho^{(A)}\otimes\ketbra{0}{0}^{\otimes N-M}$. The system $\rho^{(A)}$ is described by the $2M$ quadratures $\mathbf{R}_{\textrm{A}}$, while the ancillary modes by the $2(N-M)$ quadratures $\mathbf{R}_{\textrm{V}}$. The $N$-mode state evolves through a reservoir represented by a generic Gaussian unitary evolution $U_{\textrm{QELM}}$, whose action on the $2N$ input quadrature operators $\mathbf{R}_{\textrm{in}}=(\mathbf{R}_{\textrm{A}},\mathbf{R}_\textrm{V})$ is described by a $2N\times2N$ symplectic matrix $\mathbf{S}$ as $\mathbf{R}_{\textrm{out}}=\mathbf{S}\mathbf{R}_{\textrm{in}}$. At the output, $N$ homodyne detectors probe the vector of quadratures $\mathbf{x}_{\textrm{out}}=\mathbf{C}_{\textrm{A}}\mathbf{R}_{\textrm{A}}+\mathbf{C}_{\textrm{V}}\mathbf{R}_{\textrm{V}}$, where the matrices $\mathbf{C}_{\textrm{A}}$ and $\mathbf{C}_{\textrm{V}}$ are given by
\begin{equation}
\mathbf{C}_{\textrm{A}}=\mathbf{P}_{\theta}\mathbf{S}_{\textrm{A}},\qquad\mathbf{C}_{\textrm{V}}=\mathbf{P}_{\theta}\mathbf{S}_{\textrm{V}}. \label{eq:ca_cv_definition}
\end{equation}
Note that in Eq.~(\ref{eq:ca_cv_definition}), we write $\mathbf{S}=(\mathbf{S}_\textrm{A}\,\mathbf{S}_{\textrm{V}})$, where $\mathbf{S}_{\textrm{A}}\in \mathbb{C}^{2N\times2M}$ and $\mathbf{S}_{\textrm{V}}\in \mathbb{C}^{2N\times2(N-M)}$ represent the separate contributions of $\rho^{(A)}$ and the $N-M$ ancillary vacuum modes to the output quadratures, respectively. The matrix $\mathbf{P}_{\theta}$ is an $N\times2N$ selection matrix whose $m^{\text{th}}$ row is given by \mbox{$(0,0,\dots,\sin(\theta_m),\cos(\theta_m),\dots,0,0)$}, where the non-zero entries appear at positions $2m-1$ and $2m$. This matrix selects the $N$ output-quadrature combinations measured when the $N$ (generally unknown) local oscillator (LO) phases are $\bm{\theta}=(\theta_1,\dots,\theta_N)$. Using Eq.~(\ref{eq:ca_cv_definition}), we can compute the expectation values of the first and second moments $\langle \mathbf{x}_{\textrm{out}}\rangle$ and $\langle \mathbf{x}_{\textrm{out}}\mathbf{x}_{\textrm{out}}^T\rangle$ as 
\begin{equation}
    \langle \mathbf{x}_{\textrm{out}}\rangle=\mathbf{C}_{\textrm{A}}\langle\mathbf{R}_{\textrm{A}} \rangle,\quad \langle \mathbf{x}_{\textrm{out}}\mathbf{x}_{\textrm{out}}^T\rangle = \mathbf{C}_{\textrm{A}}\langle \mathbf{R}_{\textrm{A}}\mathbf{R}_{\textrm{A}}^T\rangle\mathbf{C}_{\textrm{A}}^T+\mathbf{N} \label{eq:output_quadratures}
\end{equation}
where $\mathbf{N}=\frac{1}{2}\mathbf{C}_{\textrm{V}}\mathbf{C}_{\textrm{V}}^{T}$ is a constant matrix accounting for the noise introduced to the output quadratures by the $N-M$ ancillary vacuum modes. Since both $\langle \mathbf{x}_{\textrm{out}}\rangle$ and $\langle \mathbf{x}_{\textrm{out}}\mathbf{x}_{\textrm{out}}^T\rangle$ depend solely on subsystem $A$, we can interpret the joint action of the reservoir and homodyne detection as an effective POVM $E(\mathbf{x})$ acting on system $A$. Specifically, $E(\mathbf{x})$ is a general Gaussian POVM induced by a Naimark dilation, which has the standard expression \cite{innocenti2023potential,zia2025quantum}
\begin{equation}
    E(\mathbf{x})=\textrm{Tr}_{\textrm{V}}(U^{\dagger}M(\mathbf{x})U(I\otimes \ketbra{0}{0}^{\otimes(N-M)})), \label{eq:naimark_dilation_POVM}
\end{equation}
where $M(\mathbf{x})=\bigotimes_{j=1}^{N}\ketbra{x_{\textrm{out},j}}{x_\textrm{out,j}}$ is the POVM associated with joint homodyne detection across the $N$ output modes. The expectation values in Eq.~(\ref{eq:output_quadratures}) can then be expressed as $\langle x_{\textrm{out},j}\rangle=\textrm{Tr}(\mathcal{M}^{(1)}_j\rho^{(A)})$ and \mbox{$\langle x_{\textrm{out},i}x_{\textrm{out},j}\rangle=\textrm{Tr}(\mathcal{M}^{(2)}_{ij}\rho^{(A)})$}, where 
\begin{equation}
\label{eq:measurement_operators}
\begin{aligned}
    \mathcal{M}_i^{(1)} &=
    \int x_{\mathrm{out},i}\,
E(\mathbf{x}_{\mathrm{out}})\,d\mathbf{x}_{\mathrm{out}}, \\
    \mathcal{M}_{ij}^{(2)} &=
    \int x_{\mathrm{out},i}x_{\mathrm{out},j}\, E(\mathbf{x}_{\mathrm{out}})\,d\mathbf{x}_{\mathrm{out}}.
\end{aligned}
\end{equation}
}

\textcolor{black}{
We then consider the observable $O$ that we aim to reconstruct via supervised learning, which we assume to be at most quadratic in the input quadratures, i.e.,
\begin{equation}
O = c\mathbf{I}+\mathbf{a}^T\mathbf{R}_{\textrm{A}}+\mathbf{R}_{\textrm{A}}\mathbf{B}\mathbf{R}_{\textrm{A}}^T, \label{eq:O_expression}
\end{equation}
where $c$ is a constant offset, $\mathbf{a}$ is a coefficient vector of length $2M$, and $\mathbf{B}$ is a $2M\times 2M$ matrix. Within the QELM framework, we wish to infer \mbox{$\langle O \rangle=y_{\textrm{tr}}$} using a linear combination of node values at the output. Due to the properties of $O$, it is natural to include both first- and second-order moments in the output feature vector $\mathbf{f}$, which we construct as \mbox{$\mathbf{f}=(1,\langle \mathbf{x}_{\textrm{out}}\rangle^T,\textrm{vec}(\langle \mathbf{x}_{\textrm{out}}\mathbf{x}_{\textrm{out}}^T \rangle)^T)^T$}. The supervised learning problem consists of finding the set of weights $\mathbf{w}$ that satisfies $y_{\textrm{tr},k} = \textrm{Tr}(O\rho_k^{\textrm{tr}})= \mathbf{f}_k^T\mathbf{w}$ for each state $\rho_k^{\textrm{tr}}$ in the training set of size $R_{\textrm{tr}}$. By forming the matrix $\mathbf{F}$ placing the $R_{\textrm{tr}}$ vectors $\mathbf{f}_k$ along its columns, the least-squares solution can be expressed in terms of the pseudoinverse $F^+$ and the vector of labels $\mathbf{y}_{\textrm{tr}}$ as $\mathbf{w}=(F^T)^{+}\mathbf{y}_{\textrm{tr}}$. As in the discrete-variable case, the set of weights identifies an operator $\mathcal{O}=((\bm{\mathcal{M}^{(1)}})^T,\textrm{vec}(\bm{\mathcal{M}}^{(2)})^T)\mathbf{w}$ that satisfies  
\begin{equation}
    \textrm{Tr}((O-\mathcal{O})\rho_{k}^{\textrm{tr}})=0 \qquad k=1,\dots,R_{\textrm{tr}}. \label{eq:O_operator_equality_CV}
\end{equation}
Therefore, $O$ and $\mathcal{O}$ have the same action on states belonging to the training set and cannot be distinguished. On the other hand, exact inference of $\mathcal{O}$ on new states $\rho^{\textrm{new}}$ requires further considerations. By hypothesis, $O$ is at most quadratic in the input quadrature operators (see Eq.(\ref{eq:O_expression})), a condition that can be written in compact form as $O=\mathbf{c}_o^T\mathbf{m}$, where $\mathbf{m}=(1, \mathbf{R}_{\textrm{A}}^T,\textrm{vec}(\mathbf{R}_{\textrm{A}}\mathbf{R}_{\textrm{A}}^T)^T)^T$ and $\mathbf{c}_o$ is a vector with components $\mathbf{c}_o=(c,\mathbf{a}^T,\textrm{vec}(\mathbf{B}^T)^T)^T$, which are uniquely identified by $O$ through Eq.~(\ref{eq:O_expression}). Similarly, from Eq.~(\ref{eq:output_quadratures}) and Eq.~(\ref{eq:measurement_operators}), we have by linearity that $\mathbf{f}=\mathbf{Z}\mathbf{m}$, where the matrix $\mathbf{Z}$ is uniquely determined by $U_{\textrm{QELM}}$ and the LO phases $\bm{\theta}$. From Eq.~(\ref{eq:O_operator_equality_CV}), it follows that
\begin{equation}
    (\mathbf{c}_o^T-\mathbf{w}^T\mathbf{Z})\mathbf{m}_k=0\quad \forall k=1,\dots,R_{\textrm{tr}}. \label{eq:informational_complete_condition}
\end{equation}
If the $R_{\textrm{tr}}$ vectors $\{\mathbf{m}_k\}_{k=1}^{R_\textrm{tr}}$ span the entire space of first- and second-order moments, i.e., $(\bm{\mathcal{M}^{(1)}},\bm{\mathcal{M}}^{(2)})$ form an informationally complete POVM set over this subspace, then Eq.~(\ref{eq:informational_complete_condition}) necessarily implies that $\mathbf{c}_o=\mathbf{Z}^T\mathbf{w}$. This condition ensures that for every new state $\rho^{\textrm{new}}$ characterized by the set of moments $\mathbf{m}_{\textrm{new}}$, we have
\begin{equation}
\label{eq:O_equality_final_VC}
\begin{split}
    \textrm{Tr}(\mathcal{O}\rho^{\textrm{new}})
    &= \mathbf{w}^T\mathbf{f}
     = \mathbf{m}_{\textrm{new}}^T\mathbf{Z}^T\mathbf{w} \\
    &= \mathbf{m}_{\textrm{new}}^T\mathbf{c}_o
     = \textrm{Tr}(O\rho^{\textrm{new}}).
\end{split}
\end{equation}
Therefore, the inference of $O$ on new test states is exact.  
The supervised learning process bypasses the need to acquire prior knowledge of the structure of the operator $O$ (described by the vector $\mathbf{c}_o$) as well as the QELM transformation and LO settings (described by the matrix $\mathbf{Z}$) by directly learning the set of weights satisfying $\mathbf{c}_o=\mathbf{Z}^T\mathbf{w}$. \\
As discussed before, the last condition is guaranteed if  the moments $\{\mathbf{m}\}_{k=1}^{R_{\textrm{tr}}}$ form a complete basis set for the first- and second-order moments. This can be achieved by training the QELM with solely coherent states and conjugated beams generated by stimulated emission (seeded parametric amplification). In this context, the use of stimulated emission is relevant because all second-order moments of independent coherent states trivially factorize, whereas quadrature correlations exist in the homodyne signals of conjugated beams. More rigorously, the number of parameters parameterizing $\langle\mathbf{R}_{\textrm{A}}\mathbf{R}_{\textrm{A}}^T\rangle$ using only $M$ independent coherent states is $2M$, which is much less than the $2M+\frac{2M(2M+1)}{2}$ parameters necessary to parametrize the most generic symmetric matrix.
Furthermore, the role of stimulated emission is pivotal for realizing rich evolutions within the QELM architecture, quadratically increasing the number of tunable parameters with the number of modes compared to passive operations.\\
Note that the use of classical training restricts the set of training states to Gaussian states, thereby limiting the available independent moments to those of first and second order. The test states, however, are not required to be Gaussian. Indeed, Eqs.~(\ref{eq:O_operator_equality_CV})--(\ref{eq:O_equality_final_VC}) make no assumptions about the nature of the test states. It is expected that including moments of order $\ge 3$ in Eq.~(\ref{eq:O_expression}) will not compromise the exact inference on new Gaussian test states, whereas performance on new non-Gaussian states will be probably limited. We leave this study for future investigation.\\
As a final consideration, we comment on the minimum number of output modes required to learn the observable $O$ satisfying Eq.~(\ref{eq:O_expression}). The condition that $\{\mathbf{m}_k\}_{k=1}^{R_\textrm{tr}}$ should span the space of first- and second-order moments implies that the $n_m\times R_{\textrm{tr}}$ matrix $(\mathbf{m}_1,\dots,\mathbf{m}_{R_\textrm{tr}})$ must have rank $n_m=1+2M+\frac{2M(2M+1)}{2}$, which is the total number of independent first- and second-order moments of an $M$-mode Gaussian state. Therefore, $R_{\textrm{tr}}\ge n_m$. The number of modes $N$ enters into the dimension of the $n_f\times R_{\textrm{tr}}$ matrix $\mathbf{F}$ through the linear inversion problem $\mathbf{y}_{\textrm{tr}}=\mathbf{F}^T\mathbf{w}$, where $n_f=1+N+\frac{N(N+1)}{2}$. The minimal setting ensuring a unique solution, with $R_{\textrm{tr}}=n_m$, is $n_f=R_{\textrm{tr}}=n_m$, which implies $\textrm{rk}(\mathbf{F})=n_m$. Solutions with $n_f<n_m$  exist provided that $\mathbf{y}_{\textrm{tr}}\in\textrm{Im}(\mathbf{F}^T)$.
}
\subsection*{Correspondence between StFWM and SpFWM\label{sec:correspondence}}
\noindent
\begin{figure}[b!]    \includegraphics[width=\linewidth]{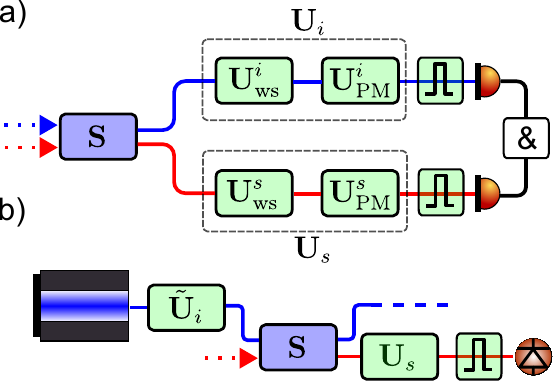}
    \caption{(a) Sequence of transformations experienced by the signal (red) and idler (blue) modes when the QELM is driven by photons generated through spontaneous FWM. Dashed lines indicate vacuum inputs.
 (b) Sequence of transformations applied to the signal and idler modes when the QELM is trained using stimulated emission.
    }
    \label{Fig:S2}
\end{figure}
Here, we first derive the expression for the coincidence probability between a signal and an idler photon at the output of the QELM in the spontaneous process. We then compare this expression with the intensity of the conjugate beam generated in the stimulated process, highlighting the conditions under which the two expressions become proportional, the essential feature exploited for training the QELM with classical signals.

We analyze both settings within the framework of input-output relations between operators. In the spontaneous process, a two-mode squeezed vacuum state is generated by SpFWM, which is characterized by the transformation matrix 
\begin{equation}
\begin{pmatrix}
\mathbf{S}_{\mathrm{ss}} &  \mathbf{S}_{\mathrm{si}} \\
\mathbf{S}_{\mathrm{is}} & \mathbf{S}_{\mathrm{ii}}
\end{pmatrix}
\label{eq:SMatrix}
\end{equation}
from the set of input signal ($\mathbf{a}_{\mathrm{s}}$) and idler ($\mathbf{a}_{\mathrm{i}}$) modes $\mathbf{a}=[\mathbf{a}_\mathrm{s},\mathbf{a}_\mathrm{i},\mathbf{a}_\mathrm{s}^{\dagger},\mathbf{a}_\mathrm{i}^{\dagger}]$. 
In Eq.(\ref{eq:SMatrix}), $\mathbf{S}_\mathrm{ss}$, $\mathbf{S}_\mathrm{ii}$ are the same-mode signal(idler) transfer functions and $\mathbf{S}_\mathrm{si}=\mathbf{S}_\mathrm{is}^{T}$ is the cross-mode transfer function related to SpFWM.
The multimode squeezed state propagates through the WS and the EOM, as illustrated in Fig.~\ref{Fig:S2}(a). The final input-output relation between operators becomes
\begin{equation}
\begin{split}
    & \mathbf{a}_{\mathrm{s}}  \rightarrow \mathbf{U}_\mathrm{s}\mathbf{S}_\mathrm{ss}\mathbf{a}_\mathrm{s}+\mathbf{U}_\mathrm{s}\mathbf{S}_\mathrm{si}\mathbf{a}_\mathrm{i}^{\dagger}\\
    & \mathbf{a}_{\mathrm{i}}  \rightarrow \mathbf{U}_\mathrm{i}\mathbf{S}_\mathrm{ii}\mathbf{a}_\mathrm{i}+\mathbf{U}_\mathrm{i}\mathbf{S}_\mathrm{is}\mathbf{a}_\mathrm{s}^{\dagger}
\end{split}
\label{eq:s_1}
\end{equation}
where $\mathbf{U}_{\mathrm{s(i)}}=\mathbf{U}_{\textup{EOM}}^{\mathrm{s(i)}}\mathbf{U}_{\textup{ws}}^{\mathrm{s(i)}}$ is the passive transformation imparted by the WS and EOM. The coincidence probability $C_{kj}$ between a signal photon in frequency-bin $k$ and an idler photon in frequency-bin $j$ is $\langle a_{\mathrm{s},k}^{\dagger}a_{\mathrm{s},k}a_{\mathrm{i},j}^{\dagger}a_{\mathrm{i},j} \rangle$, which using Eq.(\ref{eq:s_1}) becomes
\begin{equation}
    C_{kj}=\langle n_{\mathrm{s},k}\rangle\langle n_{\mathrm{i},j}\rangle+|[\mathbf{U}_\mathrm{s}\mathbf{R}\mathbf{U}_\mathrm{i}^{T}]_{kj}|^2 \label{eq:S2}
\end{equation}
where $\langle n_{\mathrm{s(i)},k(j)}\rangle = \sum_{q}|[\mathbf{U}_{\mathrm{s(i)}}\mathbf{S}_\mathrm{si(is)}]_{k(j)q}|^2$ are the average photon numbers, and $\mathbf{S}_\mathrm{ss}\mathbf{S}_\mathrm{si}=\mathbf{S}_\mathrm{si}\mathbf{S}_\mathrm{ii}^{T}=\mathbf{R}$ ensures that the transformed operators preserve the canonical commutation relations \cite{christ2013theory}. For low pair-generation probability, the first term on the right-hand side of Eq.(\ref{eq:S2}) can be neglected, $\mathbf{S}_\mathrm{ss},\mathbf{S}_\mathrm{ii}\sim \mathbf{I}$, and one recognizes $\mathbf{U}_\mathrm{s}\mathbf{S}_\mathrm{si}\mathbf{U}_\mathrm{i}^{T}$ as the low-gain biphoton wavefunction at the QELM output.

In the stimulated case, a coherent seed, here taken as the idler beam, is injected in mode $j$ at the input of the EOM and WS chain, undergoing the transformation $\tilde{\mathbf{U}}_\mathrm{i}$ shown in Fig.~\ref{Fig:S2}(b). The resulting field drives stimulated four-wave mixing in the nonlinear medium, generating a conjugate signal which subsequently undergoes the passive transformation $\mathbf{U}_\mathrm{s}$, identical to that in the spontaneous configuration. The overall transformation is
\begin{equation}
\begin{split}
    & \mathbf{a}_{\mathrm{s}}  \rightarrow \mathbf{U}_\mathrm{s}\mathbf{S}_\mathrm{ss}\mathbf{a}_\mathrm{s}+\mathbf{U}_\mathrm{s}\mathbf{S}_\mathrm{si}\tilde{\mathbf{U}}_\mathrm{i}^{\dagger}(\delta\mathbf{a}_\mathrm{i}^{\dagger}+\bm{\alpha}^*_j)\\
    & \mathbf{a}_{\mathrm{i}}  \rightarrow \mathbf{S}_\mathrm{ii}\tilde{\mathbf{U}}_\mathrm{i}(\delta\mathbf{a}_\mathrm{i}+\bm{\alpha}_j)+\mathbf{S}_\mathrm{is}\mathbf{a}_\mathrm{s}^{\dagger}
\end{split}
\label{eq:S3}
\end{equation}
where $\bm{\alpha}_j=(0,...,\alpha_j,...0)$ denotes the coherent displacement applied to the $j^{\textup{th}}$ input mode and $\mathbf{a}_{\mathrm{i}}=\bm{\alpha}_j+\delta\mathbf{a}_{\mathrm{i}}$, where $\delta\mathbf{a}_{\mathrm{i}}$ describes the noise contribution to $\mathbf{a}_{\mathrm{i}}$. The resulting signal-mode intensity $I_{kj}$ is
\begin{equation}
    I_{kj}\propto\langle n_{s,k}\rangle \sim |\alpha_j|^2|[\mathbf{U}_\mathrm{s}\mathbf{S}_\mathrm{si}\tilde{\mathbf{U}}_\mathrm{i}^*]_{kj}|^2 \label{eq:S4}
\end{equation}
where the spontaneous contribution has been neglected, being several orders of magnitude weaker than the stimulated field. By comparing Eq.(\ref{eq:S4}) with Eq.(\ref{eq:S2}), we find that $C_{kj}\propto I_{kj}$ if $\tilde{\mathbf{U}}_\mathrm{i}=\mathbf{U}_\mathrm{i}^{\dagger}$. This result is well known: if a coherent seed is prepared in the state $\mathbf{U}_\mathrm{i}^{\dagger}\bm{\alpha}_j$ and injected into the nonlinear medium of Fig.~\ref{Fig:S2}(a), then it exits the QELM in mode $j$ ($\mathbf{U}_\mathrm{i}^{\dagger}\mathbf{U}_\mathrm{i}=\mathbf{I}$), implying that it corresponds to the asymptotic output field of the system \cite{liscidini2012asymptotic}. 
\subsection*{Modeling of the QELM output}
\noindent
The states $\ket{\Psi}$ in Eqs.(\ref{eq:SP_states},\ref{eq:DP_states}) are evolved through the EOMs as $\ket{\Psi}\mapsto \mathbf{U}_{\textup{EOM}}^{s}\otimes\mathbf{U}_{\textup{EOM}}^{i}\ket{\Psi}$, where $\mathbf{U}_{\textup{EOM}}^{\mathrm{s(i)}}$ is expressed in the basis $\ket{j}$ of frequency-bins as \cite{imany2020probing,haldar2022steering}
\begin{equation}
 \mathbf{U}_{\textup{EOM}}^{\mathrm{s(i)}}  = \sum_{jn} J_n(\delta_{\mathrm{s(i)}})e^{in\theta}\ketbra{j+n}{j} \label{eq:phase_modulator}
\end{equation}
where $\delta_{\mathrm{s(i)}}$ denotes the modulation depth and $J_n$ is the $n^{\text{th}}$ Bessel function of the first kind. The relation between the amplitude of the sinusoidal driving voltage $V_{\mathrm{s(i)}}(t)=V_0\sin(\omega t+\theta_{\mathrm{s(i)}})$ and $\delta$ is $\delta = \frac{V_0}{V_{\pi}}\pi$, where $V_{\pi}$ is the half-wave voltage of the EOM. In all the numerical simulations and in the experiments, the relative phase $\theta_\mathrm{s}-\theta_\mathrm{i}$ is set to zero and $V_\mathrm{s}=V_\mathrm{i}=V_{\pi}$. The theoretically infinite basis of frequency-bins is truncated to a finite dimension of 8 and spans the logical states $\ket{-3}_{\mathrm{s(i)}},\ket{-2}_{\mathrm{s(i)}},...,\ket{3}_{\mathrm{s(i)}},\ket{4}_{\mathrm{s(i)}}$, where the convention is that the wavelength $\lambda_{s,n}$ corresponding to bin $\ket{n}_\mathrm{s}$ is $\lambda_{s,n}=1561.43+n\Delta$ and that of $\ket{n}_\mathrm{i}$ is $\lambda_{i,n}=1551.72-n\Delta$, with $\Delta=0.16$ nm. Qubit states are defined on the logical states $\ket{0}_{\mathrm{s(i)}},\ket{1}_{\mathrm{s(i)}}$, while qudit states have support on the logical states from $\ket{2}_{\mathrm{s(i)}}$ through $\ket{-1}_{\mathrm{s(i)}}$.

\subsection*{Analysis of the SNR in the stimulated and classical regime}
\label{subsection:SNR_analysis}
\begin{figure}[t!]    \includegraphics[width=\linewidth]{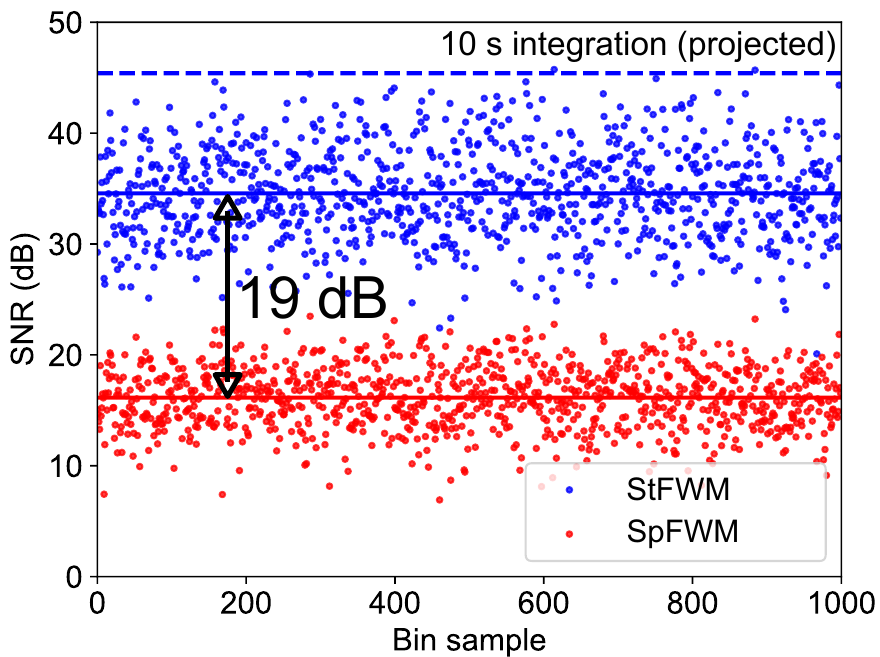}
    \caption{SNR of the brightest (see main text) frequency-bins combinations acquired by StFWM (68 ms/bin) and from SpFWM (10 s of integration), in the dual-pump configuration. The average values are indicated with straight lines. The blue dashed line indicates the projected average SNR of the StFWM acquisitions if the integration time was raised to 10 s.  
    }
    \label{Fig:S3}
\end{figure}
The main advantage of using StFWM to measure photon-pair correlations is the significant improvement in SNR compared to coincidence measurements. Indeed, the stimulated signal is proportional to the number of stimulating photons \cite{liscidini2013stimulated}, and the resulting enhancement over the pair-generation rate, for the same input pump power, has been shown to exceed five orders of magnitude in ring-resonator systems \cite{azzini2012classical}. Here, we analyze the SNR improvement of the QELM output patterns obtained via StFWM and SpFWM.
In both cases, we consider training samples from the DP configuration and sort the 36 frequency-bin combinations in descending order of intensity. We then select the bins accounting for $90\%$ of the cumulative signal, i.e., those carrying most of the signal intensity in each  $6\times 6$ output pattern. For coincidence measurements, the SNR of each bin is defined as $\frac{C_{\textrm{net}}}{\sqrt{C_{\textup{tot}}}}$, where  $C_{\textup{tot}}$ and $C_{\textrm{net}}$ denote the raw and accidentals-subtracted coincidence counts, respectively. For StFWM measurements, the SNR is defined as $\frac{I}{I_{\textrm{noise}}}$, where $I$ is the signal intensity and $I_{\textrm{noise}}\sim 1$ pW is the background noise level of the OSA (Anritsu model MS9740B, measured with acquisition settings of $0.1$ nm of resolution, $200$ Hz of video-bandwidth, $501$ sampling points, and without optical attenuator).
As shown in Fig.~\ref{Fig:S3}, the StFWM signal exhibits an average SNR of $35\pm4$ dB, whereas SpFWM yields an average SNR of $16\pm 2$ dB. This corresponds to a SNR improvement of $19(5)$ dB when using StFWM. It is important to note, however, that the two measurement schemes were performed under different acquisition conditions. Specifically, each coincidence measurement was acquired over 10 s, while the time required by the OSA to record each frequency-bin is only $\frac{34}{501}\sim 68$ ms (each frequency-bin is sampled by 34 points, and acquisition of the full 501-point spectrum takes approximately 1 s). When the two measurements are normalized to the same acquisition time, the SNR improvement increases to $19+10\log\left(\sqrt{\frac{10}{0.068}}\right)\sim 30$ dB.\\
\begin{figure*}[t!]    \includegraphics[width=\linewidth]{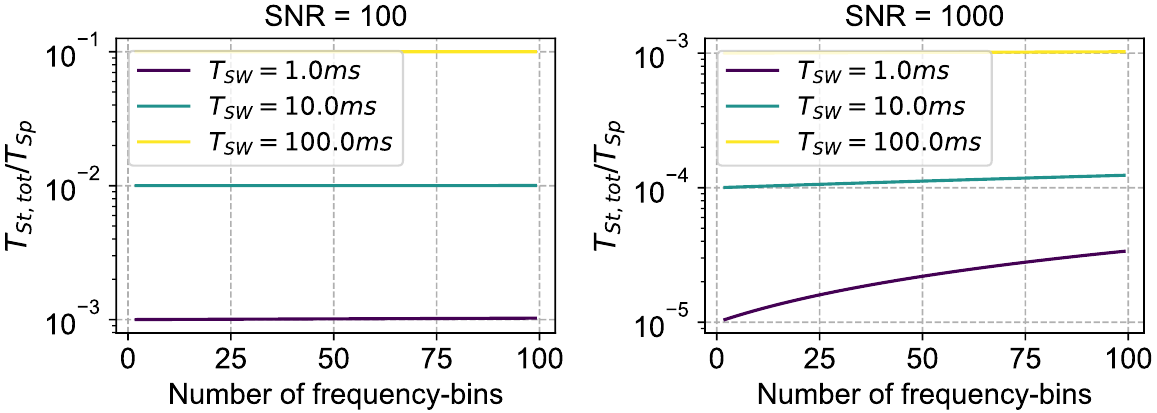}
    \caption{\textcolor{black}{Predicted ratio between the acquisition times required to achieve a target SNR using stimulated ($T_{\textrm{St,tot}}$) and spontaneous four-wave mixing ($T_{\textrm{Sp}}$) as a function of the number of frequency bins and for different seed laser switching times $T_{\textrm{SW}}$. The left and right panels correspond to SNR values of 100 and 1000, respectively. The simulation assumes that all frequency correlations can be measured simultaneously in the spontaneous process and that the same type of detector is used for both processes.}
    }
    \label{Fig:S4}
\end{figure*}
\textcolor{black}{\subsection*{Projected speed-up in integration times and analysis of the resource allocation} 
\noindent
Here we analyze the practical speed-up enabled by classical training based on stimulated emission compared with coincidence measurements in the optimal scenario where the same demultiplexing and detection apparatus is used. In this configuration, all signal-idler frequency-bin combinations can be measured simultaneously. For example, this can be achieved using only two detectors by exploiting the time-to-frequency mapping provided by a chirped Fiber Bragg grating together with time-resolved detection. The only additional overhead associated with stimulated emission is the switching time $T_{\textrm{SW}}$ required to change the wavelength of the seed laser, a step that must be repeated $N$ times for each training state, where $N$ is the local dimension (i.e., the number of frequency bins) of each photon.
For the spontaneous process, we consider a total emission bandwidth $B$ around the pump wavelength, which is divided into $N$ frequency-bin pairs with spectral width and spacing $\Delta$, such that $B=2N\Delta$. Defining the aggregate pair-generation rate over the entire bandwidth $B$ as $R_{\textrm{Sp}}$, and assuming uniform distribution over the $N^2$ signal-idler frequency-bin combinations after the QELM evolution, the number of photon pairs detected during a time interval $T_{\textrm{Sp}}$ for each frequency-bin combination is $N_{\textrm{Sp}}=\frac{R_{\textrm{Sp}}}{N^2}T_{\textrm{Sp}}$. By contrast, the number of photons detected during the stimulated process over a time interval $T_{\textrm{St}}$ is $N_{\textrm{St}}=R_{\textrm{St}}T_{\textrm{St}}$. Assuming Poissonian detection statistics, the spontaneous and stimulated processes achieve the same SNR when $N_{\textrm{Sp}}=N_{\textrm{St}}$. The stimulated process is seeded with a power $P_s$, which we assume to be equally distributed among the $N$ seed bins forming the asymptotic input field. Accordingly, we write $R_{\textrm{St}}=\bar{R}_{\textrm{St}}/N$, where $\bar{R}_{\textrm{St}}$ is the photon generation rate obtained with a monochromatic seed of power $P_s$. The same SNR is therefore achieved when $\frac{T_{\textrm{St}}}{T_{\textrm{Sp}}}=\frac{R_{\textrm{Sp}}}{\bar{R}_{\textrm{St}}N}$.
As described in the main text, the seed laser wavelength must be changed $N$ times to reconstruct the statistics of all $N^2$ frequency-bin correlations. Consequently, the total acquisition time is $T_{\textrm{St,tot}}=N(T_{\textrm{St}}+T_{\textrm{SW}})$. The ratio between the total acquisition times of the stimulated and spontaneous processes is then
\begin{equation}
\frac{T_{\textrm{St,tot}}}{T_{\textrm{Sp}}}=\frac{R_{\textrm{Sp}}}{\bar{R}_{\textrm{St}}}+N\frac{T_{\textrm{SW}}}{T_{\textrm{Sp}}}. \label{eq:time_ratio}
\end{equation}
We now express $T_{\textrm{Sp}}$ in terms of the SNR and the aggregate pair-generation rate $R_{\textrm{Sp}}$. Assuming Poissonian counting statistics, $\textrm{SNR}=\sqrt{N_{\textrm{Sp}}}=\sqrt{\frac{R_{\textrm{Sp}}T_{\textrm{Sp}}}{N^2}}$, from which we obtain $T_{\textrm{Sp}}=\frac{N^2(\textrm{SNR})^2}{R_{\textrm{Sp}}}$. Substituting this expression into Eq.~(\ref{eq:time_ratio}) gives
\begin{equation}
\frac{T_{\textrm{St,tot}}}{T_{\textrm{Sp}}}=\frac{R_{\textrm{Sp}}}{\bar{R}_{\textrm{St}}}+\frac{T_{\textrm{SW}}R_{\textrm{Sp}}}{N(\textrm{SNR})^2}. \label{eq:time_ratio_2}
\end{equation}
By expressing the aggregate spontaneous rate as \mbox{$R_{\textrm{Sp}}=NR_{\textrm{Sp}}^{\Delta}$}, where $R_{\textrm{Sp}}^{\Delta}$ is the pair-generation rate within the bandwidth $\Delta$, we obtain the final expression:
\begin{equation}
\frac{T_{\textrm{St,tot}}}{T_{\textrm{Sp}}}=\frac{R_{\textrm{Sp}}^{\Delta}}{\bar{R}_{\textrm{St}}}N+\frac{T_{\textrm{SW}}R_{\textrm{Sp}}^{\Delta}}{(\textrm{SNR})^2}. \label{eq:time_ratio_3}
\end{equation}
We now evaluate Eq.~(\ref{eq:time_ratio_3}) for realistic experimental scenarios. We consider spontaneous four-wave mixing in a standard single-mode silicon waveguide of length $L=1$ cm, nonlinear parameter \mbox{$\gamma_{\textrm{NL}}=100,\textrm{W}^{-1}\textrm{m}^{-1}$}, pumped with a power of $P_p=1$ mW, and photon pairs collected over a frequency-bin bandwidth of $\Delta=10$ GHz. The generation rate $R_{\textrm{Sp}}^{\Delta}$ is given by $R_{\textrm{Sp}}^{\Delta}=(\gamma_{\textrm{NL}}P_pL)^2\Delta=10^4\,\textrm{Hz}$ \cite{liscidini2012asymptotic}. The ratio $R_{\textrm{Sp}}^{\Delta}/\bar{R}_{\textrm{St}}$ in Eq.~(\ref{eq:time_ratio_3}) can be calculated following Ref.~\cite{helt2012does} as
\begin{equation}
\frac{R_{\textrm{Sp}}^{\Delta}}{\bar{R}_{\textrm{St}}} = \frac{\hbar\omega_p \Delta}{P_s}, \label{eq:ratio_spontaneo_stimolato}
\end{equation}
where $\omega_p$ is the angular frequency of the pump laser.
Assuming a conservative seed power of $P_s=10\,\textrm{mW}$ at a wavelength of $\lambda=1550\,\mathrm{nm}$, we obtain $R_{\textrm{Sp}}^{\Delta}/\bar{R}_{\textrm{St}} \approx 10^{-7}$. Using Eq.~(\ref{eq:time_ratio_3}), we then calculate $T_{\textrm{St,tot}}/T_{\textrm{Sp}}$ as a function of $N$ for representative values of $\textrm{SNR}=\{10^2,10^3\}$ and $T_{\textrm{SW}}=\{1,10,100\}\,\textrm{ms}$. The results are shown in Fig.~\ref{Fig:S4}.
As clearly emerges, the speed-up factor provided by StFWM is essentially independent of the number of bins and is mainly determined by the laser switching time (second term in Eq.~(\ref{eq:time_ratio_3})). For an SNR of $35$ dB, as demonstrated in the experiment, the predicted speed-up factor is on the order of $10^3-10^4$. Using the parameters discussed above, the acquisition time required to achieve an SNR of $10^3$ with StFWM is $<10$ ms, making switching times on the order of a few ms feasible.
While providing a substantial reduction in acquisition time, classical training does not introduce any additional measurement-resource overhead compared with traditional training based on coincidence measurements. Indeed, the same hardware is shared between the two approaches, and this advantage is independent of the system dimension.
}
\subsection*{Inference performance of the QELM compared to full quantum state tomography}
\label{subsection:qelm_qst_comparison}
\noindent
\begin{figure}[t!]    \includegraphics[width=\linewidth]{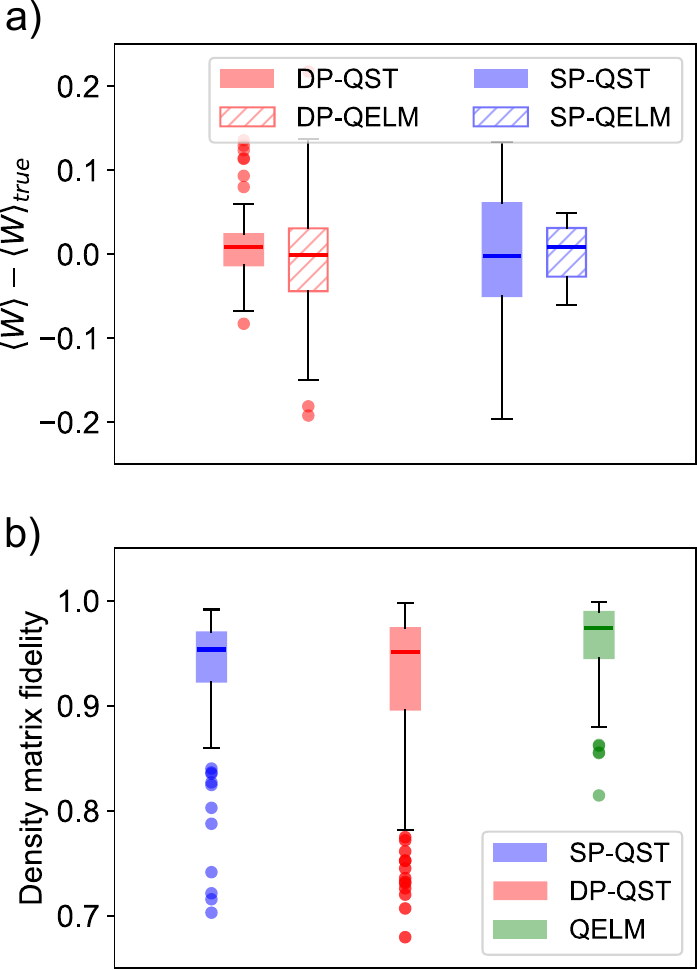}
    \caption{(a) Box plots showing the distribution of the 
    entanglement witness difference $\langle \mathcal{W}\rangle-\langle \mathcal{W}\rangle_{\textup{true}}$ in the single pump (SP) and dual pump (DP) configurations, where $\langle \mathcal{W} \rangle$ is either inferred by QELM or calculated after density matrix reconstruction (QST), and $\langle \mathcal{W} \rangle_{\textup{true}}$ is the expected value.
 (b) Box plots showing the fidelity distribution between the expected density matrices and those reconstructed by full quantum state tomography (QST) or inferred by QELM. The latter merges states obtained from single and dual-pump configurations.
    }
    \label{FIG_S5}
\end{figure}
In this section, the ability of the QELM to efficiently extract quantum features is compared with a more demanding approach based on full quantum state tomography. We first focus on entanglement witnessing in the SP and DP datasets. In Fig.~\ref{FIG_S5}(a), we report box plots showing the statistical distribution of the quantity $\langle \mathcal{W} \rangle - \langle \mathcal{W}\rangle_{\textup{true}}$ for the two datasets (DP-QELM and SP-QELM), where $\langle \mathcal{W}\rangle$ is the entanglement witness inferred by the QELM introduced in the main text, and $\langle \mathcal{W}\rangle_{\textup{true}}$ is its true value. The same distributions are also obtained by first performing full quantum state tomography (projective measurements are acquired using StFWM) to reconstruct the density matrix $\rho$ of each state, and then computing $\langle \mathcal{W}\rangle = \textrm{Tr}(\mathcal{W}\rho)$. These results (DP-QST and SP-QST) are shown as hatched box plots in Fig.~\ref{FIG_S5}(a). Interestingly, the SP-QELM distribution exhibits an interquartile range (IQR) smaller than that of SP-QST, together with a smaller 1.5-IQR range, indicating higher precision compared to SP-QST. Conversely, for the DP dataset, the entanglement witness obtained from full quantum state tomography shows a lower IQR than that inferred by the QELM. Insight into the origin of this different behavior can be gained by explicitly calculating $\langle \mathcal{W}\rangle_{\textup{SP(DP)}}$ for SP states of the form $\ket{\Psi}_{\textup{sp}}=\frac{1}{\sqrt{1+|\alpha|^2}}(\ket{00}+\alpha \ket{11} )$ and DP states of the form $\ket{\Psi}_{\textup{DP}}=\frac{1}{\sqrt{\sum |\alpha_{ij}|^2}}\sum_{ij}\alpha_{ij}\ket{ij}$, which yields
\begin{equation}
\begin{split}
\langle \mathcal{W}\rangle_{\textup{SP}} & = -\frac{|\alpha|\cos(\theta_{\alpha}) }{1+|\alpha|^2} \\
\langle \mathcal{W}\rangle_{\textup{DP}} & = \frac{|\alpha_{01}|^2+|\alpha_{10}|^2}{2}-\frac{|\alpha_{00}\alpha_{11}|\cos(\theta_{\alpha_{00}}-\theta_{\alpha_{11}})}{\sum_{ij}|\alpha_{ij}|^2} ,
\end{split}
\label{eq:witness_explicit}
\end{equation}
where $\theta_{\alpha}=\textrm{Arg}(\alpha)$ and $\theta_{\alpha_{ij}}=\textrm{Arg}(\alpha_{ij})$.

Equation (\ref{eq:witness_explicit}) shows that the witness $\langle \mathcal{W} \rangle$ is sensitive to phase errors (especially near $\theta_{\alpha}=\frac{\pi}{2}$) for SP states. While QST treats all density matrix elements with equal weight and is subject to calibration errors that accumulate over multiple projective measurements, the QELM extracts information from the most relevant density matrix elements (in this case $\bra{00}\rho\ket{00}$, $\bra{11}\rho\ket{11}$, and $\bra{00}\rho\ket{11}+\textrm{c.c}$), providing a more precise estimate of $\langle \mathcal{W} \rangle$. On the other hand, QST is more effective than the QELM for retrieving $\langle \mathcal{W} \rangle$ in the DP case, most likely because the entanglement witness depends on all the state coefficients (see Eq.(\ref{eq:witness_explicit})). Part of this information may be poorly represented or missing in the simple QELM implementation used here, which relies on a single EOM.\\
In a second analysis, we benchmark the ability of the QELM to reconstruct the density matrix of both SP and DP states against full quantum state tomography, with measurements acquired using StFWM. Figure \ref{FIG_S5}(b) shows the fidelity between the density matrices reconstructed by the two approaches (QELM and SP(DP)-QST) and their  expectations. While the median values are compatible in all cases, the use of a QELM slightly improves precision. This improvement likely arises from the fact that the QELM is constrained to learn a pure state, described by only 6 parameters, rather than the 15 free parameters spanning the full state space explored by QST. Nevertheless, the QELM reconstructs the density matrix using a single measurement setting and without calibration of the measurement apparatus, whereas QST is implemented here using 9 local Pauli measurement settings following careful calibration of the sidebands generated by each EOM.

\bibliography{Biblio}
\end{document}